\documentclass[aps,prc,twocolumn,showpacs,floatfix,nofootinbib]{revtex4} %,tightenlines,11pt,
\usepackage{graphicx}
\usepackage{dcolumn}
\usepackage{amsmath}
\usepackage[T1]{fontenc}
\usepackage{txfonts}
\usepackage{booktabs}
\usepackage[pdftex]{hyperref}
\usepackage[applemac]{inputenc}

\usepackage[usenames,dvipsnames]{color} %eks: {\color{red} nucleon??? pairs}. 

\begin{document}

\title{Evolution of the pygmy dipole resonance in Sn isotopes}

\author{H.~K.~Toft$^1$\footnote{h.k.toft@fys.uio.no}, A. C.~Larsen$^1$, A.~B\"urger$^1$, M.~Guttormsen$^1$, A. Görgen$^1$, H.~T.~Nyhus$^1$, T. Renstrøm$^1$, S.~Siem$^1$, G. M. Tveten$^1$, and A.~Voinov$^2$.\\
}

\affiliation{$^1$Department of Physics, University of Oslo, N-0316 Oslo, Norway,}
\affiliation{$^2$Department of Physics and Astronomy, Ohio University, Athens, Ohio 45701, USA}

\date{\today}

\begin{abstract}
Nuclear level density and $\gamma$-ray  strength functions of $^{121,122}$Sn below the neutron separation energy are extracted with the Oslo method using the ($^3$He,$^3$He$^\prime\gamma$)  and ($^3$He,$\alpha \gamma$) reactions.  
The level densities  of $^{121,122}$Sn display step-like structures, interpreted as  signatures of neutron pair breaking.
An enhancement in both  strength functions, compared to standard models for radiative strength, is observed in our measurements for $E_\gamma \gtrsim 5.2 $ MeV.
This enhancement is compatible with pygmy resonances centered at $\approx 8.4(1)$ and $\approx 8.6(2)$ MeV, respectively, 
and with integrated strengths corresponding to $\approx1.8^{+1}_{-5}\%$ of the  classical Thomas-Reiche-Kuhn  sum rule. 
Similar resonances were also seen in $^{116-119}$Sn. Experimental neutron-capture cross reactions are well reproduced by our pygmy resonance  predictions, while   standard strength models are less successful.
The evolution as a function of neutron number of the pygmy resonance  in $^{116-122}$Sn is described as
 a clear increase of  centroid energy  from 8.0(1)  to 8.6(2) MeV, but with no observable difference in integrated strengths. 

\end{abstract}  

\maketitle

\section{Introduction}

The level density and the $\gamma$-ray strength function are average quantities describing properties of atomic nuclei. 
The nuclear level density is defined as the  number of energy levels per unit of excitation energy, while
the $\gamma$-ray strength function may be defined as the reduced average transition probability as a function of $\gamma$-ray energy.
The strength function
characterizes average electromagnetic properties of excited nuclei. %and  has a fundamental importance for the understanding of nuclear structure and reactions involving $\gamma$-ray transitions. 

The level density and the strength function are important for many aspects of fundamental and applied nuclear physics. They are used for the calculation of cross sections and neutron-capture $(n,\gamma)$ reactions rates, which are  input parameters in, e.g., reactor physics, nuclear waste management, and  astrophysical models describing the nucleosynthesis in stars.

Tin and other heavier neutron-rich nuclei are often found display a smaller resonance for $\gamma$-ray energies below the Giant Electric Dipole Resonance (GEDR).  The existence of even a small resonance close to the neutron separation energy may have  large consequences in nuclear astrophysics on the 
calculated distribution of elemental abundance.

This article presents 
the measurements of the level densities and $\gamma$-ray strength functions in $^{121,122}$Sn  for energies  below the neutron separation energy, as well as a systematic study of the evolution of the pygmy resonances in  $^{116-119,121,122}$Sn.
The experimental results on $^{116-119}$Sn are published in Refs.~\cite{Sn_Density, Sn_Strength, 118-119}. 
All experiments  have been performed at the Oslo Cyclotron Laboratory (OCL). 

The experimental set-up is described in Sec.~II and the data analysis
in Sec.~III.  
The level densities  of $^{121,122}$Sn are presented  in Sec.~IV and the 
 strength functions in Sec.~V. 
 Section VI discusses the 
  pygmy resonance evolution and the impacts from the pygmy resonances on the ($\gamma,n$) cross sections. 
Conclusions are drawn in Sec.~VII.

\section{Experimental set-up} 

The self-supporting $^{122}$Sn target 
was enriched to $94\%$ and
had a mass thickness of  1.43~mg/cm$^{2}$. For five days the target was exposed to a 38-MeV $^3$He beam with an average current of $\approx 0.2$ nA. 
The reaction channels studied were $^{122}$Sn($^3$He,$^3$He$^\prime\gamma$)$^{122}$Sn and $^{122}$Sn($^3$He,$\alpha\gamma$)$^{121}$Sn.

Particle-$\gamma$ coincidences were recorded with 64 Si particle $\Delta E-E$ telescopes and  28 collimated NaI(Tl) $\gamma$-ray detectors. The $\Delta E$  and $E$ detector thicknesses are approximately 130~$\mu$m and 1550 $\mu$m, respectively. These detectors cover the angles of $40-54^\circ$ with respect to the beam axis, and they have a total solid-angle coverage of $\approx 9\%$ out of $4\pi$. 
The NaI detectors 
are distributed on a sphere and constitute the  CACTUS multidetector system \cite{CACTUS}.
The  detection efficiency is  $15.2\%$, and the resolution of a single NaI detector is $\approx 6\%$ FWHM,
at the $\gamma$ energy of 1332 keV. 

\section{Data analysis}
\label{sec:setup}

The  measured energy of the ejectile  is calculated into excitation energy of the residual nucleus. 
The $\gamma$-ray spectra are unfolded with the known response functions of CACTUS and by the use of the Compton subtraction method  \cite{Gut96}. The first generation $\gamma$-ray spectra are extracted  by the subtraction procedure described in Ref.~\cite{Gut87}. 

\begin{figure}[!htb]
\includegraphics[width=9cm]{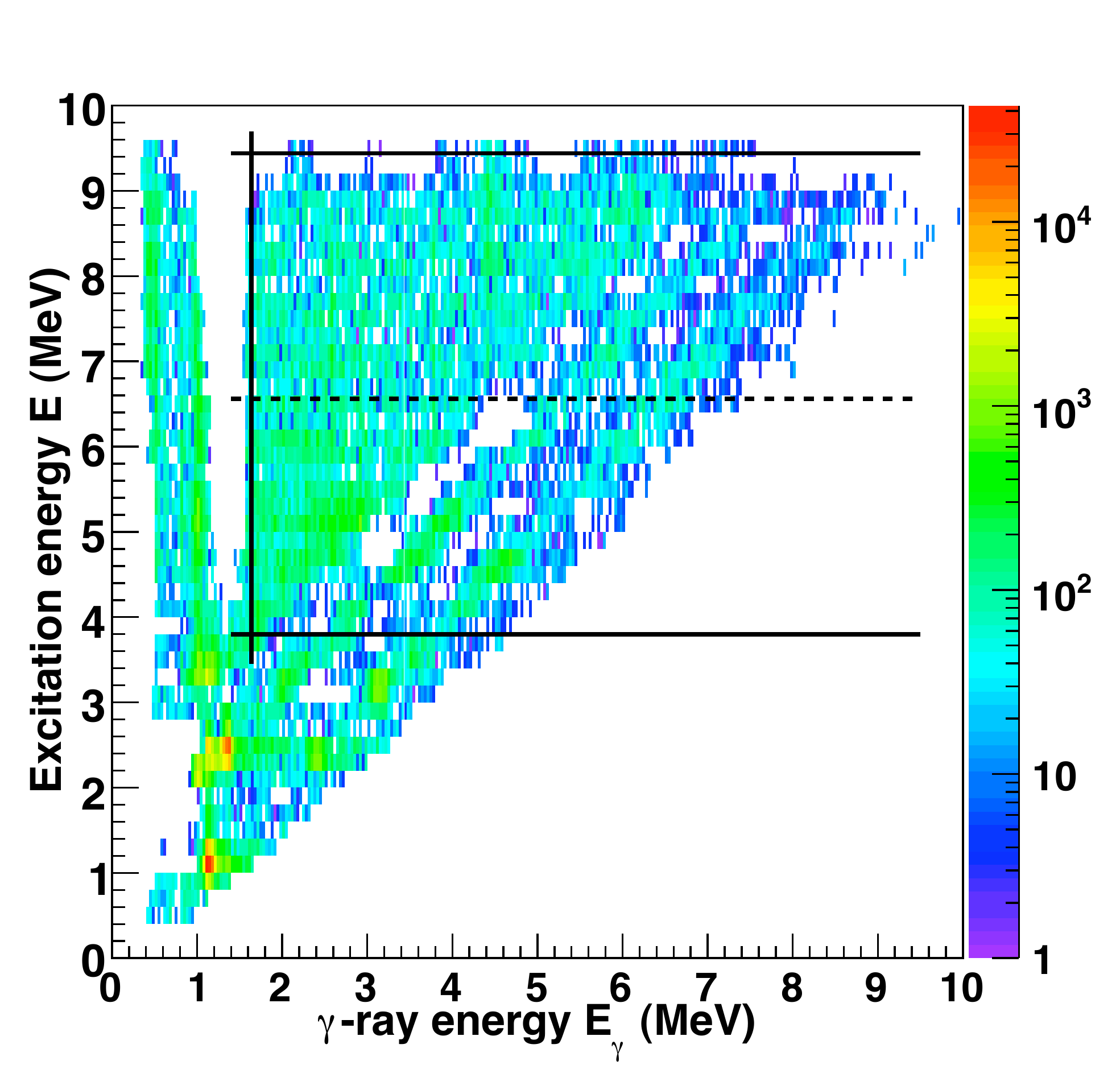}
\caption{(Color online) The first-generation matrix $P$ of $^{122}$Sn. The solid lines indicate the region for the data used in the Oslo method
($E_\gamma > 1.6$ MeV, $3.8<E<9.4$ MeV).
The dashed line ($E=6.6$ MeV)  is the middle between the upper and lower ranges in excitation energy.}
%, in a region where these dominate.}
\label{fig:fgmult122}
\end{figure}

The first-generation $\gamma$-ray spectra are arranged in a two-dimensional matrix $P(E, E_\gamma)$, shown for  $^{122}$Sn in Fig.~\ref{fig:fgmult122}.
The entries of the matrix 
give the probabilities $P(E,E_\gamma)$ that a $\gamma$-ray of energy $E_\gamma$ is emitted from a bin of excitation energy $E$.

The empty region  for low $\gamma$ energy and higher excitation energies in Fig.~\ref{fig:fgmult122} is explained by too strong subtraction 
caused by the strongly populated states (yellow/red spots) found for low $\gamma$-energy and 
lower excitation energies. Too few ﬁrst-generation $\gamma$’s remain for low $\gamma$ energy 
and higher excitation energies, which has made  the ﬁrst-generation method 
not work very well (see Ref.~\cite{Systematics}). We select and proceed with the  region between the solid lines. %in Fig.~\ref{fig:fgmult122}.
It is commented that the diagonal valleys and ridges are made up by strong  first-generation  transitions to the ground and first-excited states. %in a region where there are  few other states. 

The selected region of the first generation matrix $P$ 
is factorized into the level density $\rho$ and the radiative transmission coefficient ${\cal T}$  \cite{Sch00a}:
\begin{equation}
\label{probab}
P(E,E_\gamma)\propto \rho(E-E_\gamma) {\cal T}(E_\gamma).
\end{equation}
The factorization into two independent components is justified for nuclear reactions leading to a compound state prior to a subsequent $\gamma$ decay \cite{Bohr-Mottelson}. 
The factorization is performed by an iterative procedure \cite{Sch00a} where the independent functions $\rho$ and ${\cal T}$ are adjusted until a global $\chi^2$ minimum with  the experimental $P(E,E_\gamma)$ is reached.
 
\begin{figure*}[!bt]
\includegraphics[width=19cm]{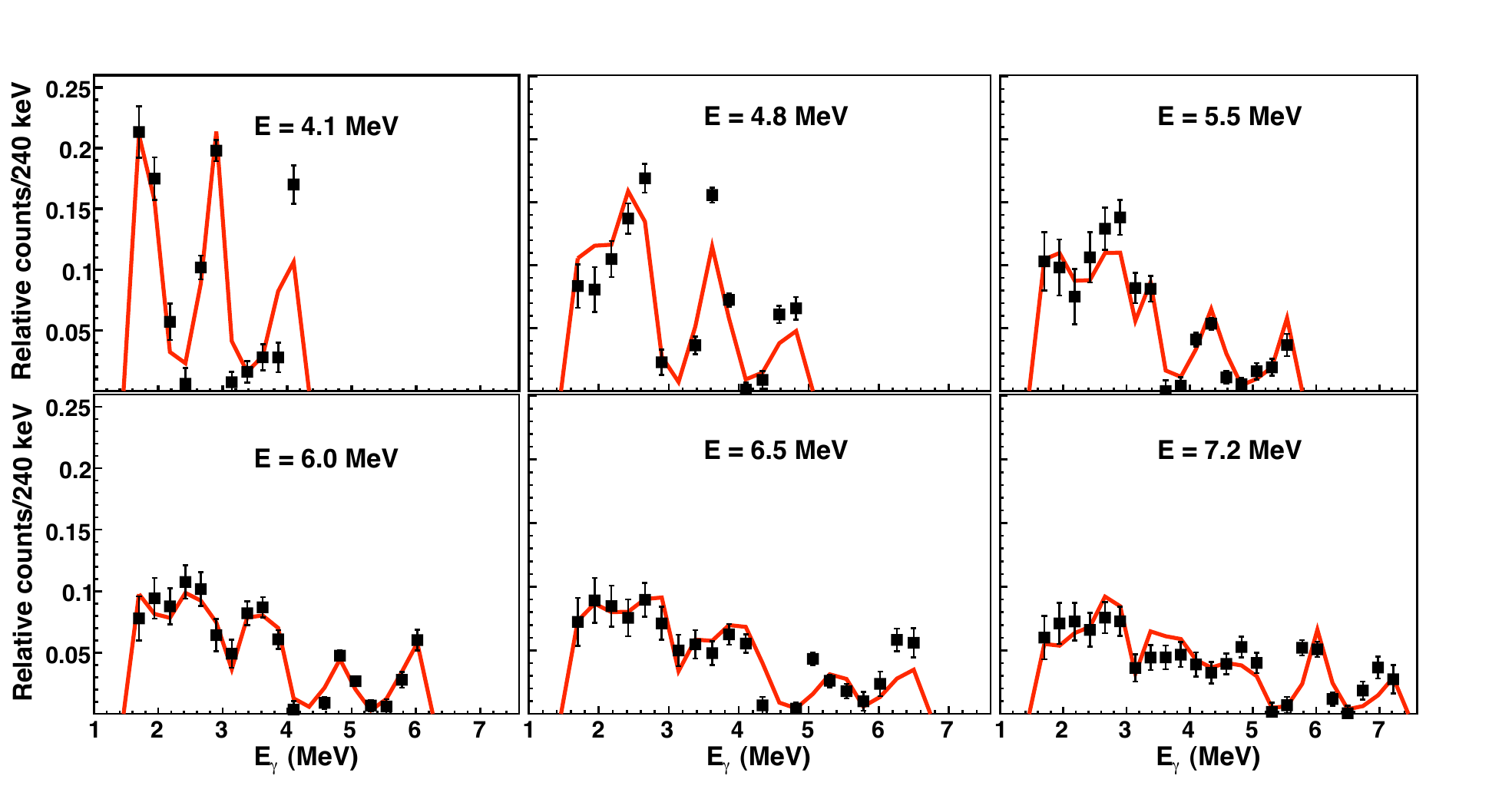}
\caption{(Color online) Comparison between experimental (squares) and  $\chi^2$ fitted (solid lines) $P$ matrix  for  $^{122}$Sn. 
 The energy bins have been compressed to 240 keV/ch in $E$ and in $E_\gamma$.
While the panels show the results for the indicated example excitation energies, the  fit has been  performed globally for the entire region of the $P$ matrix that has been selected for the analysis  (see Fig.~\ref{fig:fgmult122}).} 
\label{does-it-work}
\end{figure*}

 The quality of  the factorization of level density and strength function is illustrated for $^{122}$Sn  in Fig.~\ref{does-it-work}.
At example excitation energies (indicated on the panels), the entries of the $P$ matrix obtained from the  $\chi^2$-fitted output functions $\rho$ and $\cal{T}$ using Eq.~(\ref{probab}) are compared to those of the  experimental  $P$ matrix.
The fitted output (solid line) agrees well with experimental data. It is noted that in some of the panels, the fitted curves are significantly lower than the experimental  data points (For $E=4.1$ MeV: the transition to the ground state;   for $E=4.8$ MeV: the transition to the first-excited state). 
This is probably explained by the fit adjusting the entire matrix, and not just these example excitation energies. 

 The Brink-Axel hypothesis~\cite{Bri55,Axe62}  states that 
 the GEDR and any other collective excitation mode built
on excited states have the same properties as those built on
the ground state. 
Equation \ref{probab} shows that the 
 transmission coefficient is assumed to be independent of excitation energy $E$, which is a consequence of the Brink-Axel hypothesis. 

\begin{figure}[!htb]
\includegraphics[width=9.5cm]{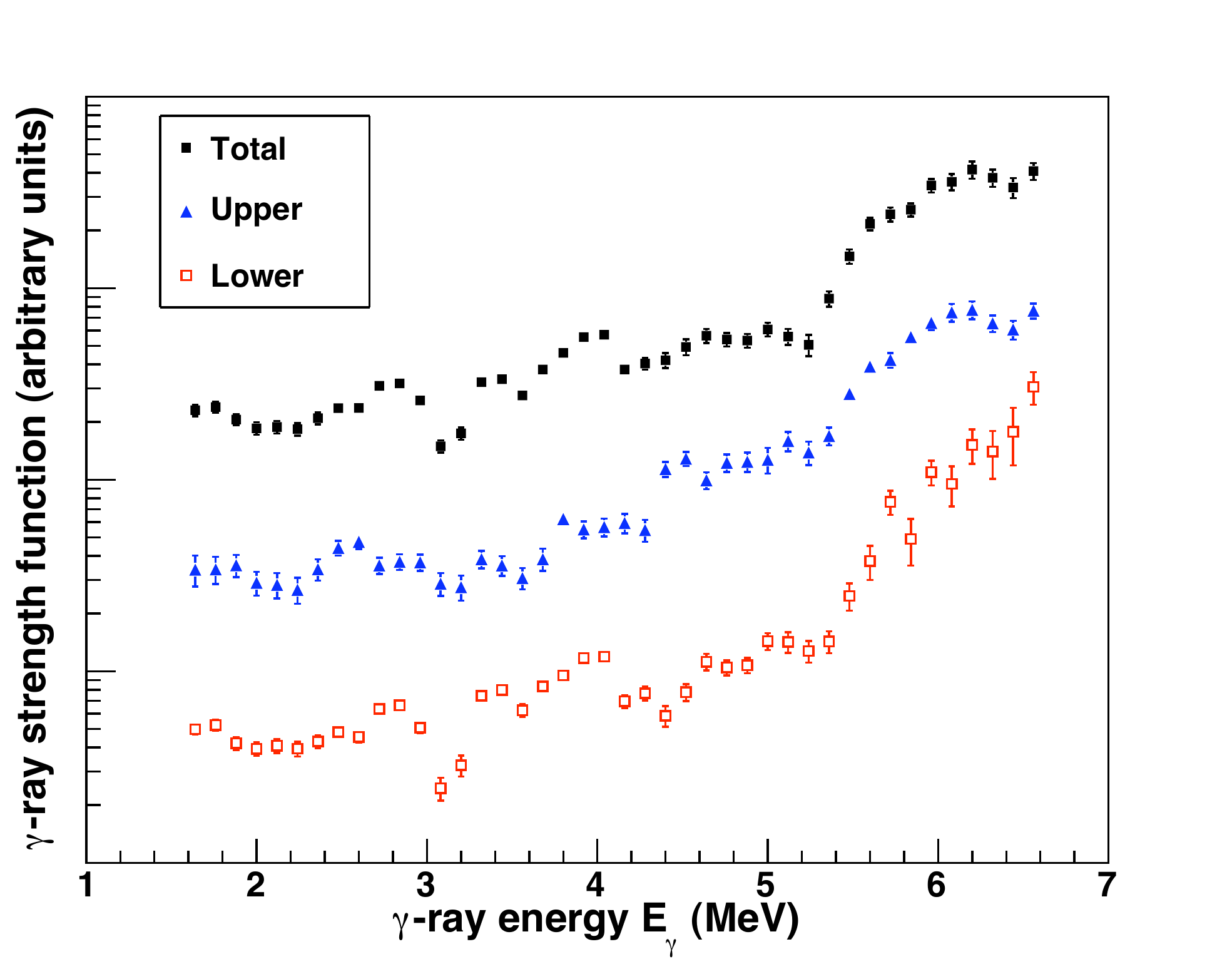}
\caption{(Color online) Comparison of unnormalized $\gamma$-ray strength functions (arbitrary units) for  $^{122}$Sn. The shown strength functions are derived from the independent data sets of the upper and lower part of the selected region of the $P$ matrix, as well as from the total selected region for comparison.}
%from the total selected region of the $P$ matrix as well as  from the independent data sets of the upper and lower part.} %The   total strength function is shown for comparison.}
\label{fig:strength-ulike-122}
\end{figure}

Figure \ref{fig:strength-ulike-122} shows an investigation of this assumption for $^{122}$Sn, which is of special concern due to some clear  structures in the strength function.
We divide the selected region of the $P$ matrix into two parts (separated by the dashed line in Fig.~\ref{fig:fgmult122}), which are two independent data sets. 
Figure \ref{fig:strength-ulike-122} displays the  
 strength functions derived from the lower and upper  part, as well as from the  total region. The strength functions,
 proportional to ${\cal T}/{E_\gamma}^3$, are not normalized and are shown in arbitrary units.
As the clear structures are found at the same locations for the two independent data sets,
 the ${\cal T}$ is indeed found to be approximately independent of excitation energy.

The adjustment to Eq.~(\ref{probab}) determines only the  functional forms of $\rho$ and $\cal T$. These two functions are invariant under the following transformations \cite{Sch00a}:
\begin{eqnarray}
\tilde{\rho}(E-E_\gamma)&=&A\exp\left[\alpha\left(E-E_\gamma\right)\right]\,\rho(E-E_\gamma)\,,
\label{eq:array1}\\
\tilde{\cal T}(E_\gamma)&=&B\exp\left(\alpha E_\gamma\right) \,{\cal T}(E_\gamma)\,.
\label{eq:array2}
\end{eqnarray}
The parameters $A$ and $B$ define the correction to the absolute values of the functions $\rho$ and ${\cal T}$, respectively, while the parameter $\alpha$ defines  their common correction to the log-scale slope. These parameters will be determined in Secs.~IV and V.

\section{Level densities}
\label{LD-results}

The constants $A$ and $\alpha$ needed to normalize the experimental level density $\rho$, 
are determined using literature values of the  known discrete energy levels at low energy  and of  the level  spacing $D$ at the neutron separation energy $S_n$.  
We use the same normalization procedure as in Refs.~\cite{Sn_Density, Sn_Strength, 118-119}, in order to have a common ground for comparison. 

The chosen model is the back-shifted Fermi gas (BSFG) model, published by von Egidy {\em et al.} in 1988 \cite{Egi88}. 
The level density at the neutron separation energy $\rho(S_n)$ is calculated from known values of the $s$-wave level spacing $D_0$  \cite{Sch00a}:
 \begin{eqnarray}
 \label{eq:D0}
\rho(S_n) &=& \frac{2{\sigma}^2}{D_0}\cdot \left\{\left(I_t + 1\right)\exp\left[ \frac{-\left(I_t +1\right)^2}{2\sigma^2}\right]  \right. \nonumber \\
  &+& \left. I_t  \exp\left[ \frac{-{I_t}^2}{2\sigma^2}\right] \right\}^{-1} \,,
 \end{eqnarray}
where $I_t$ is the target spin, and where the spin-cutoff parameter $\sigma$  is also evaluated at $S_n$. The spin-cutoff parameter is defined as $\sigma^{2}~=~0.0888 \, A^{2/3} aT$, where $A$ is the mass number of the isotope, and $T$ is the nuclear temperature given by $T = \sqrt{U/a}$. Here,  $U$ is the nucleus intrinsic excitation energy, and $a$ is the level-density parameter.  The parameterization used for $a$ is $a = 0.21 \, A^{0.87}~{\rm MeV^{-1}}$. The parameterization of $U$ is $U = E - E_{\rm pair} - C_1$, where  the pairing energy $E_{\rm pair}$ is calculated from the proton and neutron pair-gap parameters: $E_{\rm pair} = \Delta_{\rm p} +\Delta_{\rm n}$, and where the back-shift parameter $C_1$ is defined as $C_1 = -6.6 \, A^{-0.32}$.  

The experimental value of  $D_0$ for $^{121}$Sn is found in Ref.~\cite{RIPL-3} 
and is used to calculate $\rho(S_n)$ using the input parameters listed in Tab.~\ref{tab:parametre}. The pair-gap parameters are evaluated from even-odd mass differences \cite{Wapstra} according to the method of Ref.~\cite{Dob01}. 

\begin{table*}[!htb] 
\caption{Input parameters and the resulting values for the calculation of the normalization value $\rho(S_n)$.}
\begin{tabular}{|l|cccccccc|c|c|}
\hline
\hline
Nucleus     & $S_n$  &  $D_0$ & $a$    & $C_1$   & $\Delta_{\rm n}$ & $\Delta_{\rm p}$    & $\sigma(S_n)$  & $\rho$($S_n$) & $\eta$ \\ 
          & (MeV)  &  (eV) &  (MeV$^{-1}$)  & (MeV)   & (MeV) & (MeV)   & &  $(10^{4}$ MeV$^{-1})$  & \\
\hline
$^{121}$Sn  & 6.17   &   1250(200)  &   13.62  & -1.42 & 0           & 0.82 &  4.57   & 3.42(86)  &   0.25  \\
$^{122}$Sn  &  8.81  &    62(31)$^a$        &  13.72  & -1.42  &   1.32   &  1.12  &    4.75  &  20(10)$^a$  &  0.42 \\
\hline
\hline
\end{tabular}
\label{tab:parametre} \\
$^{a}$ Estimated from systematics.
\end{table*}

\begin{figure}[!ht]
\includegraphics[width=9.5cm]{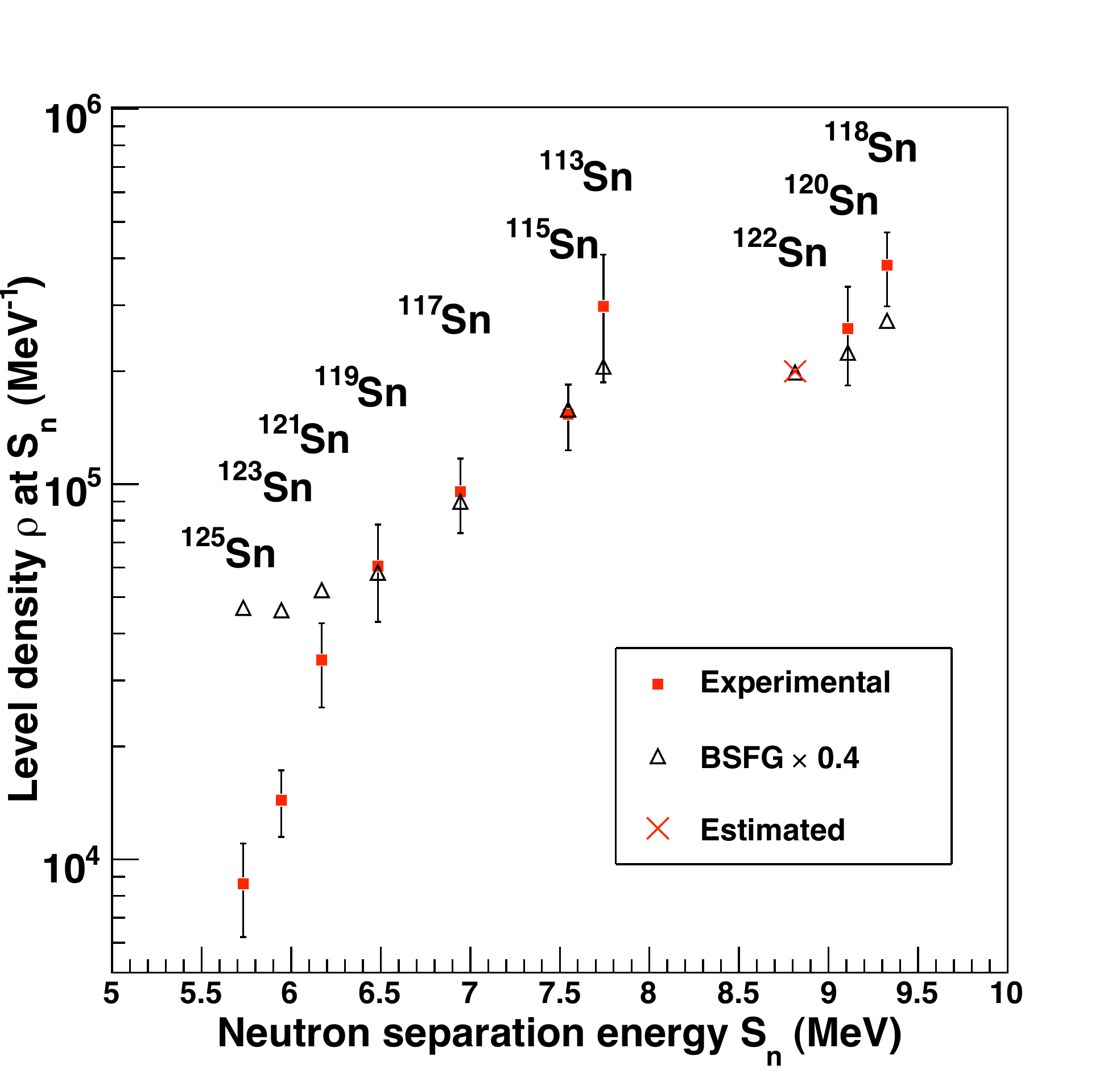}
\caption{(Color online) Estimation (cross) of the experimental value of $\rho(S_n)$ for  $^{122}$Sn from systematics. Experimental values (squares) and global BSFG predictions (triangles)  for  $\rho(S_n)$  are shown for various Sn isotopes as function of  $S_n$. (See text.)}
\label{Rho-syst}
\end{figure}

No experimental value exists for $D_0$ of  $^{122}$Sn, and we estimate $\rho(S_n)$ for this isotope  from systematics.  Figure \ref{Rho-syst} shows 
$\rho(S_n)$  calculated from the experimental values of $D_0$ according to Eq.~(\ref{eq:D0})
for all available Sn isotopes   as a function of $S_n$.
The  values of  $D_0$ have been taken from
 Ref.~\cite{RIPL-3}. We have also calculated  $\rho(S_n)$ according to the prediction of the BSFG model  \cite{Gil65}:
\begin{equation}
\label{eqn:rhoBSFG}
\rho(E)_{\rm BSFG} =\frac{\mathrm{exp}\left(2\sqrt{aU}\right)}{12\sqrt{2}\,a^{1/4}\,U^{5/4}\,\sigma}\,,
\end{equation}
with the above-listed parameterizations. %\cite{Egi88}.
The theoretical value of $\rho(S_n)$, multiplied with a common factor of 0.4,
are shown in Fig.~\ref{Rho-syst}  together with the experimental values. From the trends appearing  in the figure, we estimate $\rho(S_n)$ for $^{122}$Sn to $2.0(10)\cdot 10^5 \,{\rm MeV}^{-1}$ (50\% uncertainty assumed, see Tab.~\ref{tab:parametre}).  

While we would like to normalize to  $\rho(S_n)$, our experimental data only cover the excitation energy region from 0 to  $S_n - 2$ MeV, due to methodical limitations. 
We therefore make an interpolation from our measurements to  $S_n$ using  the BSFG prediction in Eq.~(\ref{eqn:rhoBSFG}).
The prediction is multiplied by a scaling parameter $\eta$ (see Tab.~\ref{tab:parametre}) in order to agree with the normalization value $\rho(S_n)$:
\begin{equation}
\rho(E)_{\rm BSFG}\rightarrow\eta\,\rho(E)_{\rm BSFG}\,.
\end{equation}

\begin{figure}[!ht]
\includegraphics[width=9.5cm]{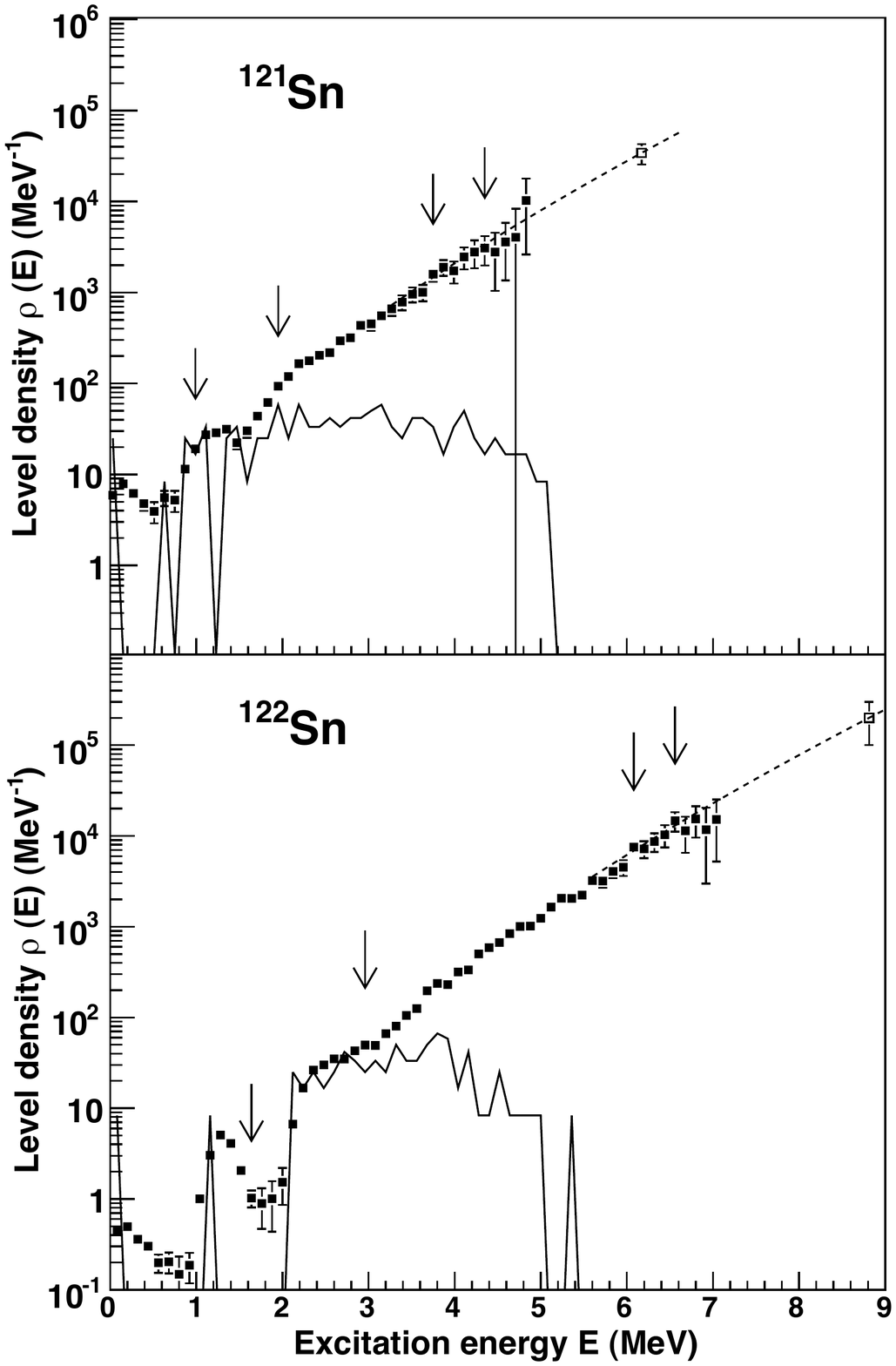}
\caption{Normalized level densities of $^{121,122}$Sn (filled squares) as a function of excitation energy with energy bins of 120 keV/ch. 
The solid lines represent the discrete level densities obtained from spectroscopy. 
The dashed line in both panels is the BSFG prediction, which  is used for interpolation, scaled with $\eta$ to coincide with  $\rho(S_n)$ (open square). The value of   $\rho(S_n)$  has been calculated from neutron resonance data.   
The arrows indicate the two regions used for normalization.}
\label{rhonorm}
\end{figure}

Figure \ref{rhonorm} shows the normalized level densities of $^{121,122}$Sn. 
 The arrows  indicate the two regions that have been used for normalization to the discrete level density and to the normalization value $\rho(S_n)$.  
As expected, the level density function of $^{121}$Sn is very similar to those of the other even-odd nuclei $^{117,119}$Sn, while the level density function of  $^{122}$Sn is very similar to those of the even-even nuclei $^{116,118}$Sn  \cite{Sn_Density,118-119}. 

The level densities of $^{121,122}$Sn in Fig.~\ref{rhonorm} show  step-like structures, a feature also seen in $^{116-119}$Sn \cite{Sn_Density,118-119}. 
In $^{121,122}$Sn, two pronounced bumps are seen in the region of $\approx 0.8-1.4$ MeV and  $\approx 1.8-2.4$ MeV. The corresponding steps are located at $\approx 1.0$ and $\approx 2.0$ MeV, respectively. The second step is
very abrupt, especially  in the even-even nucleus, and the step is followed by  a significantly higher level density.
The second step is therefore a candidate for the neutron pair-breaking process in $^{121,122}$Sn. 
Such neutron pair-breaking bumps are especially distinctive in Sn isotopes since the proton number  is magic  (Z = 50), making  proton pair breaking occur only at relatively higher excitation energies.  
A detailed discussion of the pair-breaking process has been given in Refs.~\cite{Sn_Density,118-119}.

\section{Gamma-ray strength functions}
\label{sub-norm}
\label{section-gamma}

The  $\gamma$-ray transmission coefficient ${\cal T}$, which is deduced from the experimental data, is related to the
$\gamma$-ray strength function $f$  by
\begin{equation}
{\cal T}(E_\gamma) = 2\pi \sum_{XL}E_\gamma^{2L+1}f_{XL}(E_\gamma)\,,
\end{equation}
where $X$ denotes the electromagnetic character and $L$ the multipolarity of the  $\gamma$ ray.
The transmission coefficient ${\cal T}$  is normalized in log-scale slope  ($\alpha$)  and in absolute value ($B$) (see Eq.~(\ref{eq:array2})). 

For  $s$-wave neutron resonances and assuming a major contribution from dipole radiation and
 parity symmetry for all excitation energies,
 the  expression for the average radiative width $\left<  \Gamma_\gamma (E, I, \pi)\right>$   will at $S_n$ reduce to  \cite{Voinov01}:
\begin{eqnarray}
\label{eq:Gammagamma}
\lefteqn { \left<  \Gamma_\gamma (S_n, I_t \pm 1/2,\pi_t)\right> } \nonumber \\
& = & \frac {B}{4\pi\,\rho(S_n, I_t \pm 1/2, \pi_t )} \int_0^{S_n} {\rm d}E_\gamma \, {\cal T}(E_\gamma)\,\rho(S_n-E_\gamma) \nonumber \\
& & \times \sum_{J = -1}^{1}g (S_n - E_\gamma, I_t \pm 1/2 + J)\,.
\end{eqnarray}
Here, $I_t$ and $\pi_t$ are the spin and parity of the target nucleus in the neutron capture $(n,\gamma)$ reaction.
We determine $B$ by using  the BSFG model for the  spin distribution $g$ given in Ref.~\cite{Egi88} and the  experimental value of 
$\left<\Gamma_\gamma(S_n)\right>$.

\begin{figure}[!ht]
\includegraphics[width=9.5cm]{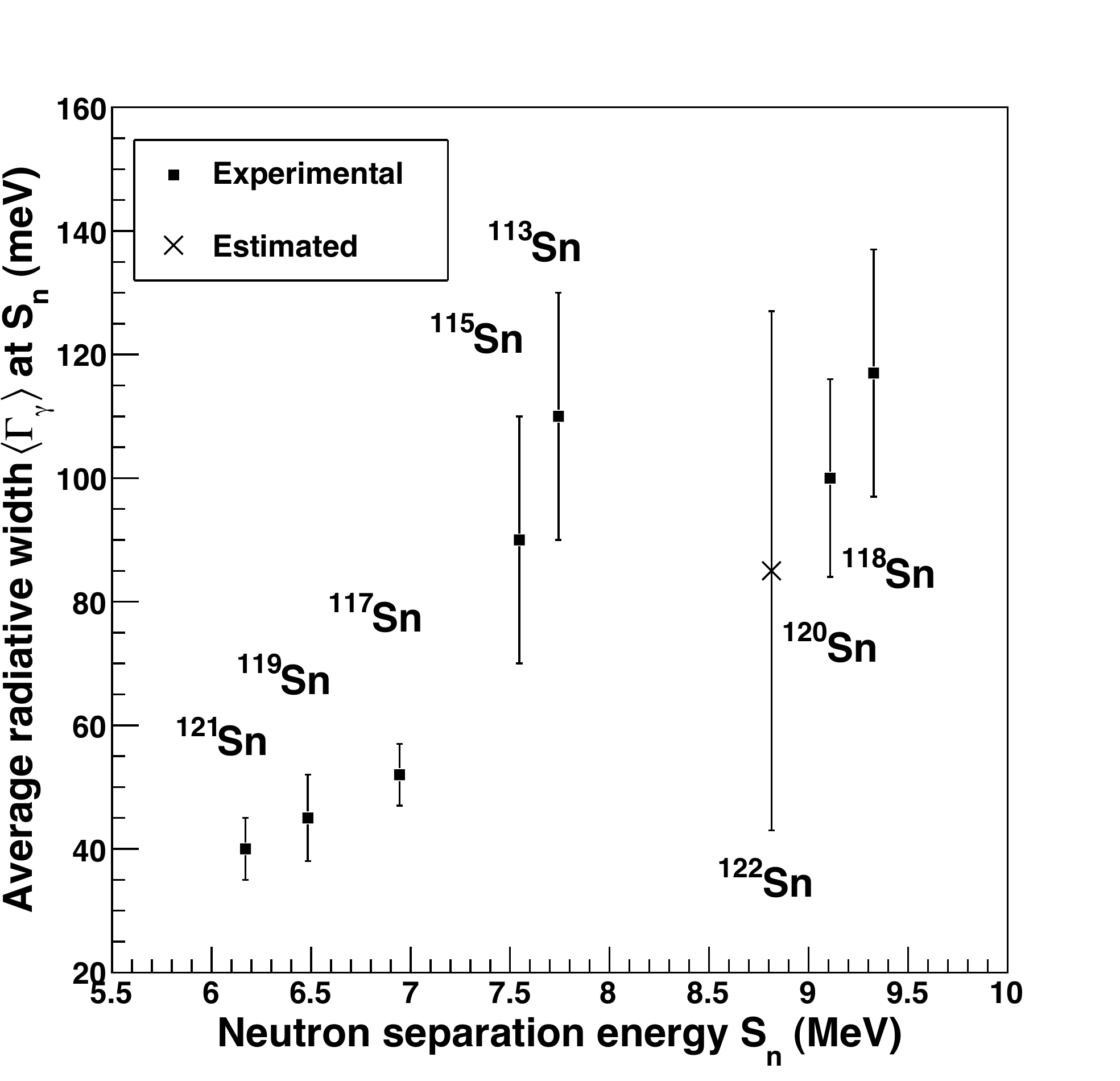}
\caption{Estimation of the average radiative width at $S_n$,  $\left<\Gamma_\gamma(S_n)\right>$, for $^{122}$Sn. The respective values of other Sn isotopes are plotted (squares) as a function of   $S_n$. (See text.)}
\label{Gamma-syst}
\end{figure}

For $^{121}$Sn, the 
 radiative width at the neutron separation energy is  available in literature. For  
 $^{122}$Sn, we  estimated it from systematics. Figure \ref{Gamma-syst} shows  the $\left<  \Gamma_\gamma(S_n)\right>$  plotted against $S_n$ for   Sn isotopes where this quantity is known (taken from Ref.~\cite{RIPL-3}).
From the appearing trend of the even-even nuclei, we estimate $\left<  \Gamma_\gamma(S_n)\right>$ to $85(42) \,{\rm meV}$ for $^{122}$Sn. 
The applied input parameters needed for  determining the normalization constant $B$ for $^{121,122}$Sn  are  shown in Tab.~\ref{tab:styrke-normering}. The values for $^{121}$Sn have been taken from Ref.~\cite{RIPL-3}.

\begin{table}[!htb] 
\caption{Input parameters for normalization of the $\gamma$-ray transmission coefficient ${\cal T}$  of $^{121,122}$Sn.}
\begin{tabular}{|l|ccc|}
\hline
\hline
Nucleus   & $I_t$ & $D_0$ & $\left<\Gamma_{\gamma}(S_n)\right>$  \\ 
       &  ($\hbar$) & (eV) & (meV) \\
\hline
$^{121}$Sn   &  0  & 1250(200)  & 40(5)  \\
$^{122}$Sn   &  3/2 & $62(31)^a$ & $85(42)^a$ \\
\hline
\hline
\end{tabular}\\
$^a$ Estimated from systematics.
\label{tab:styrke-normering}
\end{table}

\begin{figure}[!htb]
\includegraphics[width=9.5cm]{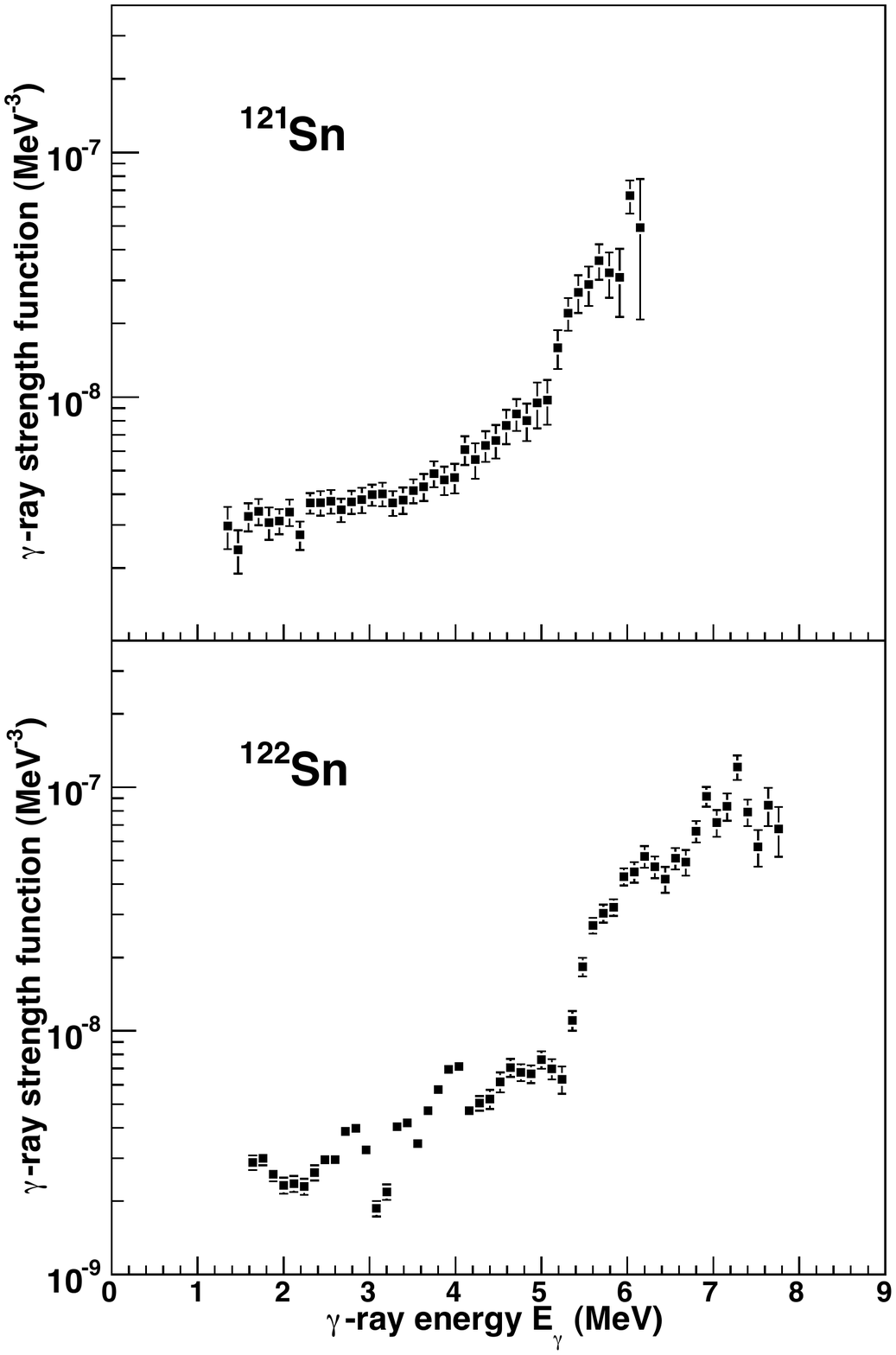}
\caption{Normalized $\gamma$-ray strength functions of $^{121,122}$Sn as functions of $\gamma$-ray energy. The energy bins  are 120 keV/ch.}
\label{fig:strength_both}
\end{figure}

The normalized $\gamma$-ray strength functions of  $^{121,122}$Sn are shown in Fig.~\ref{fig:strength_both}.  
The error bars show the statistical uncertainties.
While the strength function of $^{121}$Sn is smooth, just like those of $^{116-119}$Sn \cite{Sn_Strength, 118-119},  the strength function of $^{122}$Sn  displays clear structures in the entire $E_\gamma$ region. As discussed in Sec.~III,
these structures also appear using different, independent data sets.

\begin{figure*}[tb]
\includegraphics[width=19cm]{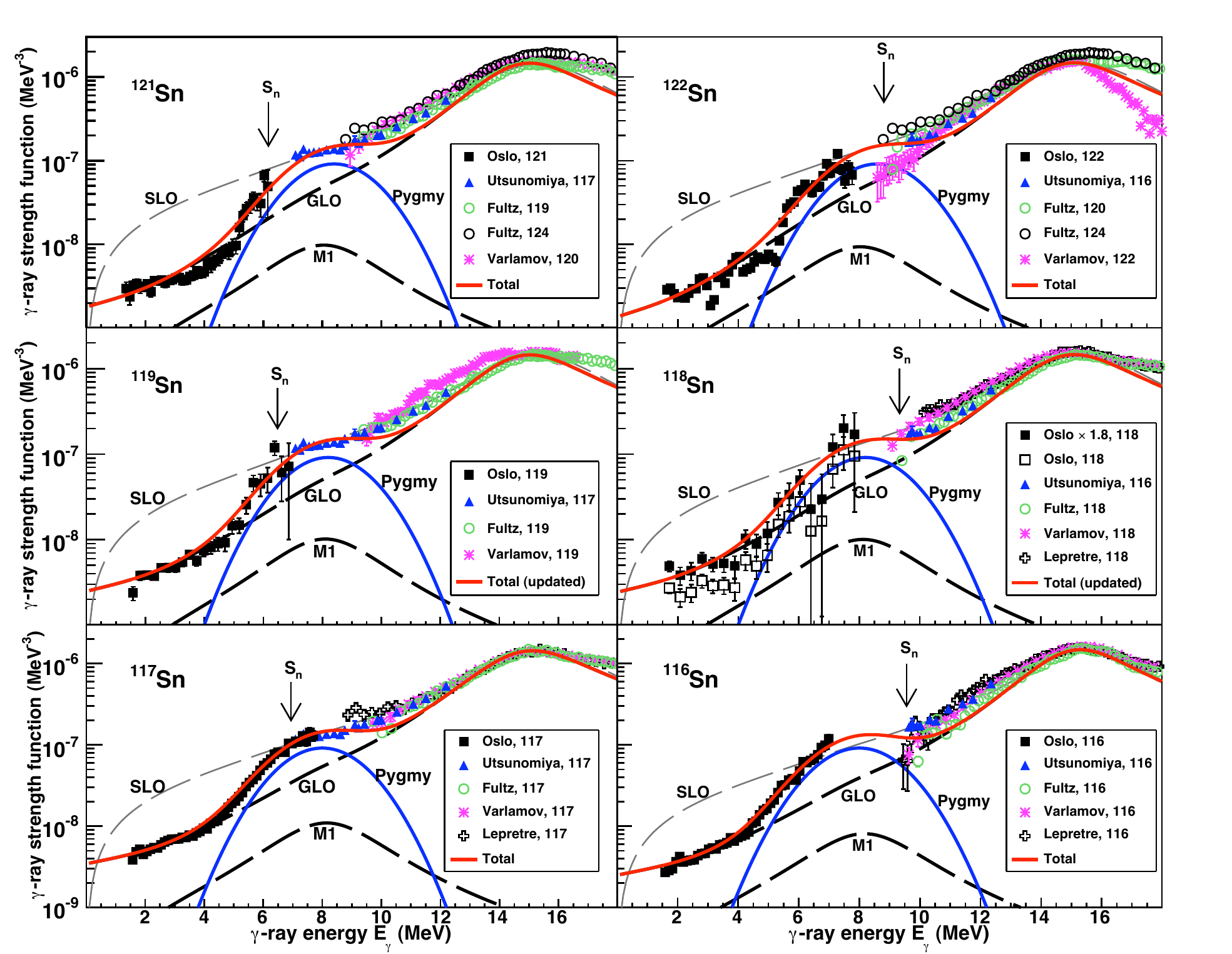}
\caption{(Color online) Comparison of  the prediction of the total strength functions with  the OCL experimental measurements for  $^{116-119,121,122}$Sn. 
The strength function  of $^{118}$Sn is shown both as multiplied with 1.8 (filled squares) and as the original normalization (open squares). The arrows indicate $S_n$. 
The total strength predictions (solid lines) are modeled as Gaussian pygmy resonance additions to the GLO (E1 + M1) baselines. The SLO (E1 + M1) baselines are also shown.
\\
Upper, left panel: 
$^{117}{\rm Sn}(\gamma,n)$ from Utsunomiya {\em et al.}~\cite{Utsunomiya}, $^{119,124}{\rm Sn}(\gamma,x)$ from Fultz {\em et al.}~\cite{Fultz69}, and $^{120}{\rm Sn}(\gamma,x)$ from Varlamov {\em et al.}~\cite{Varlamov03}.
\\
Upper, right panel:  $^{116}{\rm Sn}(\gamma,n)$ \cite{Utsunomiya},  $^{120,124}{\rm Sn}(\gamma,x)$ \cite{Fultz69}, and $^{122}{\rm Sn}(\gamma,n)$ from Varlamov {\em et al.}~\cite{Varlamov09}.
\\
Middle, left panel: $^{117}{\rm Sn}(\gamma,n)$ \cite{Utsunomiya}, $^{119}{\rm Sn}(\gamma,x)$ \cite{Fultz69}, and $^{119}{\rm Sn}(\gamma,n)$ \cite{Varlamov09}.
\\
Middle, right panel: $^{116}{\rm Sn}(\gamma,n)$ \cite{Utsunomiya},  $^{118}{\rm Sn}(\gamma,x)$ \cite{Fultz69}, $^{118}{\rm Sn}(\gamma,x)$ from Varlamov {\em et al.}~\cite{Varlamov03}, and $^{118}{\rm Sn}(\gamma,x)$ from Lepr\^{e}tre {\em et al.}~\cite{Lepretre74}.
\\
Lower, left panel: 
$^{117}{\rm Sn}(\gamma,n)$ \cite{Utsunomiya},  $^{117}{\rm Sn}(\gamma,x)$ \cite{Fultz69}, $^{117}{\rm Sn}(\gamma,x)$ \cite{Varlamov03}, and $^{117}{\rm Sn}(\gamma,x)$ \cite{Lepretre74}. 
\\
Lower, right panel: $^{116}{\rm Sn}(\gamma,n)$ \cite{Utsunomiya},  $^{116}{\rm Sn}(\gamma,x)$ \cite{Fultz69}, $^{116}{\rm Sn}(\gamma,x)$ \cite{Varlamov03}, and $^{116}{\rm Sn}(\gamma,x)$ \cite{Lepretre74}.
}
\label{fig:pygme}
\end{figure*}

A change of the log-scale slope in the strength functions,  leading to a sudden increase of strength, is seen in  $^{121,122}$Sn for $E_\gamma > 5.2$ MeV. The change of log-scale slope represents the onset of a small resonance, commonly related to as the pygmy dipole resonance. A comparison of our $^{121,122}$Sn measurements compared with photonuclear cross section data from Refs.~\cite{Utsunomiya, Varlamov09, Fultz69, Varlamov03, Lepretre74}
is shown 
in the two upper panels
in Fig.~\ref{fig:pygme}. 
Similar strength increases were also seen in $^{116-119}$Sn \cite{118-119}, and this 
figure will be further discussed for those isotopes when discussing the evolution of the pygmy resonance in the next section.

In order to investigate the experimental strength functions of $^{121,122}$Sn, 
we have applied commonly used models for the GEDR resonance and for the magnetic spin-flip resonance, also known as the Giant Magnetic Dipole Resonance (GMDR).

The Generalized Lorentzian (GLO) model \cite{Kopecky87} is used for the GEDR resonance. The GLO  model is known to  agree rather well both for  low $\gamma$-ray energies and for the GEDR centroid 
at about 15 MeV. The strength function approaching a non-zero value for low $E_\gamma$ 
is not a property specific for the Sn isotopes, but has  been seen for all nuclei studied at the OCL so far. 

In the GLO model, the $E1$ strength function is given by  \cite{Kopecky87}:
\begin{eqnarray}
f^{\rm GLO}_{E1} (E_\gamma)  &=&  \frac{1}{3\pi^2\hbar^2 c^2}\sigma_{E1}\,\Gamma_{E1} \nonumber \\
& \times &\left[ E_\gamma \,\frac{\Gamma_{KMF}(E_\gamma,T_f)}{\left({E_\gamma}^2 - {E_{E1}}^2\right)^2 + {E_\gamma}^2 \left(\Gamma_{KMF}(E_\gamma,T_f)\right)^2} \right. \nonumber \\
  &+& \left. 0.7 \, \frac{\Gamma_{KMF}(E_\gamma=0, T_f)}{{E_{E1}}^3}\right] \,,  \nonumber \\
 \end{eqnarray}
in units of ${\rm MeV}^{-3}$, 
where the Lorentzian parameters are the GEDR's centroid energy $E_{E1}$, width $\Gamma_{E1}$ and cross section $\sigma_{E1}.$ These experimental parameters are not available for $^{121,122}$Sn. We instead apply  
 the one measured for  $^{120}$Sn to  $^{121}$Sn, and the one measured for   $^{124}$Sn to $^{122}$Sn, from Fultz  \cite{Fultz69}
   (see Tab.~\ref{tab:parametre-teori}).

\begin{table}[!htb] 
\caption{Applied parameters for the parameterization of the GEDR and the GMDR contributions for $^{121,122}$Sn.} 
\begin{tabular}{lccccccc}
\hline
\hline
Nucleus   & $E_{E1}$ &  $\Gamma_{E1}$ & $\sigma_{E1} $ & $E_{M1}$ & $\Gamma_{M1}$ & $\sigma_{M1}$  & $T_f$ \\ 
       &  (MeV) &  (MeV) & (mb)& (MeV) &  (MeV) & (mb) & (MeV) \\
\hline
$^{121}$Sn  & 15.53  & 4.81  & 253.0  & 8.29 & 4.00 & 1.11  & 0.25(5) \\
$^{122}$Sn  & 15.59  &  4.77  &  256.0 &  8.27 & 4.00 & 1.09   &  0.25(5) \\
\hline
\hline
\end{tabular}
\label{tab:parametre-teori}
\end{table}

The GLO model is  
temperature dependent from the incorporation of a temperature dependent width $\Gamma_{KMF}$. This width    is   the term responsible  for    the non-vanishing $E1$ strength at low excitation energy. It has been adopted from the Kadmenski{\u{\i}}, Markushev and Furman (KMF) model \cite{KMF} and is given by:
\begin{equation}
\Gamma_{KMF}(E_\gamma, T_f )= \frac{\Gamma_r}{{E_r}^2}\left( {E_\gamma}^2 + 4\pi^2 {T_f}^2 \right)\,,
\end{equation}
in units of MeV, and where $T_f$ is the temperature. 

Usually,  $T_f$  is interpreted as the nuclear temperature of the final state, with the commonly applied expression $T_f=~\sqrt{U/a}$. 
In this work and in Refs.~\cite{Sn_Density,Sn_Strength,118-119}, we  assume a constant temperature, i.e., the $\gamma$-ray strength function is  
independent of excitation energy. This approach  is adopted for consistency with the Brink-Axel hypothesis (see Sec.~\ref{sec:setup}). %, which leads to  the assumption of a temperature-independent strength function.

Moreover, we treat $T_f$  as a free parameter. This is necessary to adjust the theoretical strength prediction to   our   low-energy measurements. 
The applied values of the $T_f$ parameters are listed in Tab.~\ref{tab:parametre-teori}. 

The $M1$ spin-flip resonance is modeled with the functional form of  a Standard Lorentzian (SLO)  model \cite{RIPL-2}:
\begin{equation}
f_{M1}^{\rm SLO}(E_\gamma) = \frac{1}{3\pi^2\hbar^2 c^2} \frac{\sigma_{M1} {{\Gamma}_{M1}}^2 E_\gamma}{\left({E_\gamma}^2 - {E_{M1}}^2\right)^2 + {E_\gamma}^2 \, {\Gamma_{M1}}^2} \,,
\end{equation}
where the parameter $E_{M1}$ is the centroid energy, $\Gamma_{M1}$ the width and $\sigma_{M1}$ the cross section of the GMDR. These Lorentzian parameters are for $^{121,122}$Sn predicted from the theoretical expressions in Ref.~\cite{RIPL-2} and shown  in Tab.~\ref{tab:parametre-teori}. 
The predictions for the GEDR using the GLO model and for the GMDR for $^{116-119,121,122}$Sn nuclei are shown as dashed lines in Fig.~\ref{fig:pygme}.

The  Standard Lorentzian (SLO) model was also tested 
and is included in Fig.~\ref{fig:pygme} (the $M1$ spin-flip resonance contribution is added to it). The SLO succeeds in reproducing the $(\gamma,x)$ data, but clearly fails for the low-energy strength measurements, both when it comes  to   absolute value and  shape. The same has been the case also for many other nuclei measured at the OCL and elsewhere. Therefore, we consider the SLO not to be adequate below the neutron threshold. 

At present, it is unclear how these resonances should be modeled properly, although many theoretical predictions exist. 
We have found \cite{Sn_Strength,118-119} that the Sn pygmy resonance is satisfactorily   reproduced 
by a Gaussian distribution \cite{Sn_Strength}:
\begin{equation}
f_{\rm pyg}(E_\gamma) = C_{\rm pyg} \cdot \frac{1}{\sqrt{2\pi} \, \sigma_{\rm pyg}} \exp\left[ \frac{-\left( E_\gamma - E_{\rm pyg}\right)^2}{2{\sigma_{\rm pyg}}^2}\right) \,,
\end{equation}
superimposed on the GLO prediction.
Here,  $C_{\rm pyg}$ is the resonance's absolute value normalization constant, $E_{\rm pyg}$ the centroid energy and  $\sigma_{\rm pyg}$  the width.  These parameters are treated as  free.  

By adding the discussed theoretical strength contributions, the prediction of the total 
 $\gamma$-ray strength function is given by:
\begin{equation}
f_{\rm total} = f_{E1} + f_{M1} +  f_{\rm pyg}\,.
\end{equation}

We determined the Gaussian pygmy parameters of $^{121,122}$Sn from fitting to our measurements. The centroid energies of the pygmy resonances are 8.4(1) and 8.6(2) MeV, respectively. We found that the same width $\sigma_{\rm pyg}$ and strength $C_{\rm pyg}$ as in  $^{116,117}$Sn \cite{118-119} gave a very good agreement also in $^{121,122}$Sn, so the width and strength are kept unchanged. The pygmy parameters are listed in Tab.~\ref{tab:pygme}. The estimated error bars given in the table take into account systematic uncertainties in the normalization  values and in the choice of baseline of the  pygmy resonance, including the fact that the GLO does not perfectly follow  the $(\gamma,n)$ measurements for higher $E_\gamma$ values. 

\begin{table}[!htb] 
\caption{Empirical values of  Gaussian pygmy parameters, and the corresponding integrated strengths and TRK values of the pygmy resonances, in  $^{121,122}$Sn.}
\begin{tabular}{lccccc}
\hline
\hline
Nucleus    & $E_{\rm pyg}$  & $\sigma_{\rm pyg }$ & $C_{\rm pyg}$ & Integrated strength &TRK value  \\ 
        & (MeV) & (MeV)  & ($10^{-7}{\rm MeV}^{-2})$ & (MeV$\cdot$mb) & (\%) \\
\hline
$^{121}$Sn  &  8.4(1) & 1.4(1)  & $3.2^{+3}_{-9}$   & $31^{+3}_{-8}$ & $1.8^{+1}_{-5}$  \\
$^{122}$Sn  &  8.6(2) & 1.4(1)  &  $3.2^{+3}_{-9}$  &  $32^{+2}_{-9}$  & $1.8^{+1}_{-5}$    \\
\hline
\hline
\end{tabular}
\label{tab:pygme}
\end{table}

The predictions for $^{121,122}$Sn  are shown as solid lines in the upper panels of Fig.~\ref{fig:pygme}.
We see that the predictions of the total  strength describe the measurements rather well, in the sense that  the Gaussian pygmy resonances fill a very large fraction 
of the missing strength.
Still,  the  Gaussian distribution does not completely  follow neither the left flank nor the right flank of the pygmy resonances in $^{121,122}$Sn. 
In the case of  $^{116,117}$Sn \cite{Sn_Strength,118-119},  having  a larger $T_f$ of $T_f = 0.46(1)$ MeV, the left flank was followed very well. However, in all Sn isotopes, there is a gap on the right flank between measured data and the GLO.  A better pygmy resonance representation than the Gaussian or a better model for the baseline  may be found in the future.

Strength from the  resonances in $^{121,122}$Sn have been added in the energy region of $\approx 5 - 8$ MeV according to our measurements, and in the region of $\approx 5 - 11$ MeV when comparing to photonuclear data as well.
 The total  integrated strengths of the pygmy resonances, based on the total predictions, are estimated to $32_{-9}^{+3}$ ~MeV$\cdot$mb. This constitutes $1.8_{-5}^{+1}\%$ of the classical Thomas-Reiche-Kuhn (TRK) sum rule, assuming all pygmy resonance strength being $E1$, see Tab.~\ref{tab:pygme}. 

Even though uncertainties in the choice of baseline have been considered in the uncertainty estimates, another  prediction of the GEDR than the GLO or another function for the pygmy resonance than the Gaussian may be found in the future. This will   consequently  influence the estimates on the pygmy resonance.  Lack of data, i.e., the gaps between our measurements and the  $(\gamma,n)$ measurements in resonance  region, and also the lack of    $(\gamma,n)$ measurements  for $^{121,122}$Sn, adds to the uncertainties in the estimates of the pygmy resonances.

%Several factors not considered for the uncertainty estimates also influence the accuracy of the pygmy predictions: Lack of data at the centroid of the resonance for all  isotopes except $^{117}$Sn,
% choice of baseline,  the GLO not perfectly following the measurements, choice of function for the pygmy prediction, lack of GEDR parameters, lack of ($\gamma,n$) measurements, and choice of normalization of our measurements.

Measurements from other reactions and using other methods have also been used to estimate the TRK value of the Sn pygmy, and these estimates deviate from each other. 
Data from $^{116,117}$Sn($\gamma$,n) experiments  \cite{Utsunomiya} indicate an integrated strength which constitutes $\approx 1\%$ of the TRK sum rule, which agrees within the uncertainty with our value.
From   $^{116,124}$Sn($\gamma$,$\gamma^\prime$) experiments \cite{Govaert}, the TRK value is calculated to  $0.4-0.6\%$.
This may seem to deviate from our result. However, taking into account unresolved strength  in the quasi-continuum of typically a factor of $2-3$, the ($\gamma$,$\gamma^\prime$) results are compatible within the uncertainty with  the other data.

\section{Evolution of the pygmy resonance}

\begin{figure}[!htb]
\includegraphics[width=9.5cm]{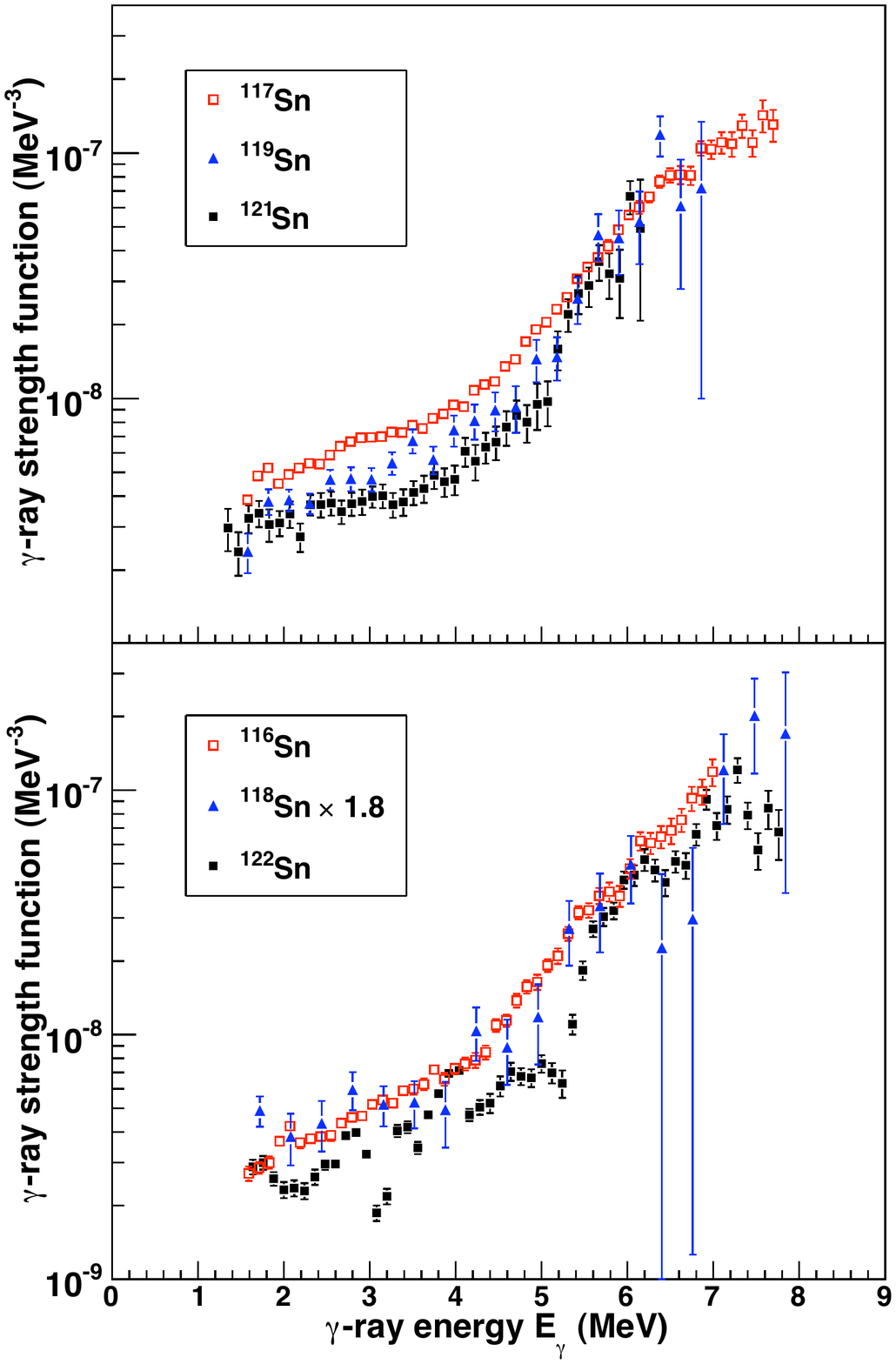}
\caption{(Color online) Normalized $\gamma$-ray strength functions as functions of $\gamma$-ray energy for the Sn isotopes measured at the OCL. The even-odd isotopes, $^{117,119,121}$Sn, are shown in the upper panel, while the even-even isotopes, $^{116,118,122}$Sn,  are shown in the lower.
The measured strength of $^{118}$Sn has been multiplied by 1.8 (see Ref.~\cite{118-119}).
The energy bins are 120 keV/ch for $^{116,117,121,122}$Sn, 240 keV/ch for $^{119}$Sn and 360 keV/ch for $^{118}$Sn.}
\label{fig:strength_both-alle}
\end{figure}

Studying the  neutron dependency is important and  may help in determining the origin of the Sn pygmy resonance.
Figure \ref{fig:strength_both-alle} shows the present and previously analyzed normalized strength functions for the Sn isotopes. 
The measurements of  $^{118}$Sn have been multiplied by 1.8 in order to agree with those of $^{116}$Sn (see Ref.~\cite{118-119}).

First, it may seem like  a trend that the tail of the strength function decreases in strength as the neutron number $N$ increases. 
Second,  it is noticeable from Fig.~\ref{fig:strength_both-alle} that  the change of log-scale slope, which  represents the onset of the pygmy resonance, 
 occurs at an higher $E_\gamma$ value in  $^{121,122}$Sn than for $^{116,117}$Sn. The changes of slope are clearest for the even-even nuclei. They   are found at
$\approx 4.5$ MeV in $^{116}$Sn and at $\approx 5.2$ MeV in $^{122}$Sn.

The values of $T_f$ for $^{121,122}$Sn are lower than for $^{116-119}$Sn. 
There is no physical reason for different nuclear temperatures. Lowering the parameter $T_f$ is instead necessary in order to get the lowest-energy part of the GLO comparable in magnitude with the  measurements. 

The centroid energy $E_{\rm pyg}$ of the pygmy resonances in $^{121,122}$Sn has larger values than those of earlier studies in  $^{116,117}$Sn  \cite{Sn_Strength, 118-119}. For $^{121,122}$Sn, the pygmy centroids ire 8.4(1) and 8.6(2) MeV (see Tab.~\ref{tab:pygme}), respectively, while 8.0(1) MeV is found for $^{116,117}$Sn \cite{Sn_Strength, 118-119}.  
During the pygmy resonance fitting, it was clear that the centroid energies had to be significantly increased for the heavier isotopes.  
The signiﬁcant  increases are also apparent  from studying the energies for 
which there is a change of log-scale slope in the strength functions. %The change of slope  represent the onset of the pygmy resonance. 
%This agrees well with the finding of an increase of the localization of the log-scale slopes  in the heavier isotopes compared to the lighter.  
Moreover, keeping the centroid energy constant has as a consequence  
that the same pygmy width $\sigma_{\rm pyg}$ and pygmy strength $C_{\rm pyg}$ as in $^{116,117}$Sn \cite{118-119} also give the best fit   in $^{121,122}$Sn.

In the earlier study of $^{118,119}$Sn \cite{118-119}, the data have large error bars. This means that the pattern of an increasing centroid energy  was not so clear, and the choice then was to keep the centroid energy  constant while compensating with an   increase of the resonance width.
We have updated the resonance prediction of  $^{118,119}$Sn by following the same pattern. The estimated centroid energy of the pygmies in $^{118,119}$Sn is then 8.2(1) MeV, while the width and strength  are kept constant. 
Updated parameter values are listed in   Tab.~\ref{tab:pygme116-119} and displayed in Fig.~\ref{fig:pygme}. The parameters  for  the GEDR and  GMDR contributions are listed in Tab.~\ref{tab:parametre-teori116-119}.

\begin{table}[!tb] 
\caption{Empirical values of $^{116-119}$Sn   Gaussian pygmy parameters, and the corresponding pygmies' integrated strengths and TRK values. 
For $^{118}$Sn, the values have been found from fitting to the measured strength function multiplied by 1.8.}
\begin{tabular}{lccccc}
\hline
\hline
Nucleus    & $E_{\rm pyg}$  & $\sigma_{\rm pyg }$ & $C_{\rm pyg}$ & Integrated strength &TRK value  \\ 
        & (MeV) & (MeV)  & ($10^{-7}{\rm MeV}^{-2})$ & (MeV$\cdot$mb) & (\%) \\
\hline
$^{116}$Sn  &  8.0(1) & 1.4(1) & $3.2^{+3}_{-9}$  &  $30^{+0}_{-8}$ & $1.7^{+0}_{-4}$ \\
$^{117}$Sn  &  8.0(1) & 1.4(1)  & $3.2^{+3}_{-9}$ &  $30^{+0}_{-8}$ & $1.7^{+0}_{-4}$ \\
$^{118}$Sn  &  8.2(1) & 1.4(1)  & $3.2^{+0}_{-9}$  &  $30^{+0}_{-8}$ & $1.8^{+0}_{-5}$ \\
$^{119}$Sn  &  8.2(1) & 1.4(1)  & $3.2^{+0}_{-9}$  &  $30^{+0}_{-8}$ & $1.7^{+0}_{-4}$ \\
\hline
\hline
\end{tabular}
\label{tab:pygme116-119}
\end{table}

\begin{table}[!tb] 
\caption{Applied parameters for the parameterization of the GEDR and the GMDR contributions for $^{116-119}$Sn.} 
\begin{tabular}{lccccccc}
\hline
\hline
Nucleus   & $E_{E1}$ &  $\Gamma_{E1}$ & $\sigma_{E1} $ & $E_{M1}$ & $\Gamma_{M1}$ & $\sigma_{M1}$  & $T_f$ \\ 
       &  (MeV) &  (MeV) & (mb)& (MeV) &  (MeV) & (mb) & (MeV) \\
\hline
$^{116}$Sn  &  15.68 &  4.19 & 266.0 &  8.41 & 4.00 & 0.773  & 0.46(1) \\
$^{117}$Sn  & 15.66 & 5.02   & 254.0 &  8.38 & 4.00 & 1.04 &  0.46(1)  \\
$^{118}$Sn   &  15.59 &  4.77 & 256.0 & 8.36 & 4.00 & 0.956  & 0.40(1) \\
$^{119}$Sn  &  15.53 &  4.81 & 253.0 & 8.34 & 4.00 & 0.963 &  0.40(1)  \\
\hline
\hline
\end{tabular}
\label{tab:parametre-teori116-119}
\end{table}

\begin{figure}[bt]
 \begin{center}
 \includegraphics[width=9.5cm]{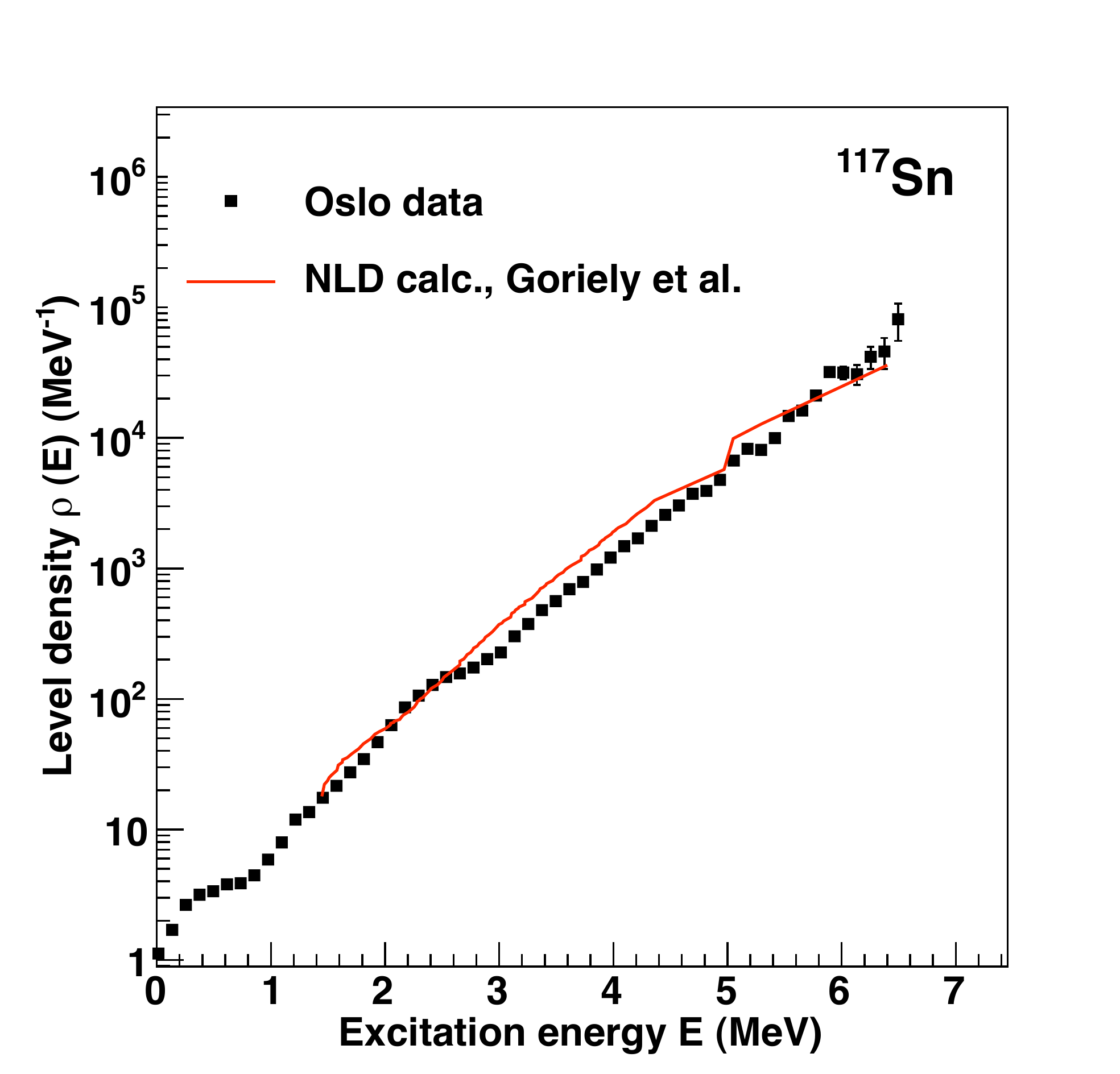} %[clip,width=\columnwidth]
 \caption{(Color online) Level-density measurements on $^{117}$Sn compared with microscopic calculations 
 from Ref.~\cite{go08}.}
 \label{fig:gorNLD}
 \end{center}
 \end{figure}

We would like  to investigate for several isotopes the effect  of our predicted pygmy resonances on the $(n,\gamma)$  cross sections and compare these with  available experimental  measurements.
This has been done for  $^{117-119,121}$Sn
using the reaction code TALYS~\cite{TALYS}. For the level density, we have applied the spin- and parity-dependent 
calculations of Goriely, Hilaire and Koning~\cite{go08}, which  are in good agreement with our level-density data,
as demonstrated for $^{117}$Sn in Fig.~\ref{fig:gorNLD}. Also, we have used  the neutron optical potential of Koning and Delaroche \cite{arjan}. 

\begin{figure*}[bt]
 \begin{center}
 \includegraphics[height=13.5cm]{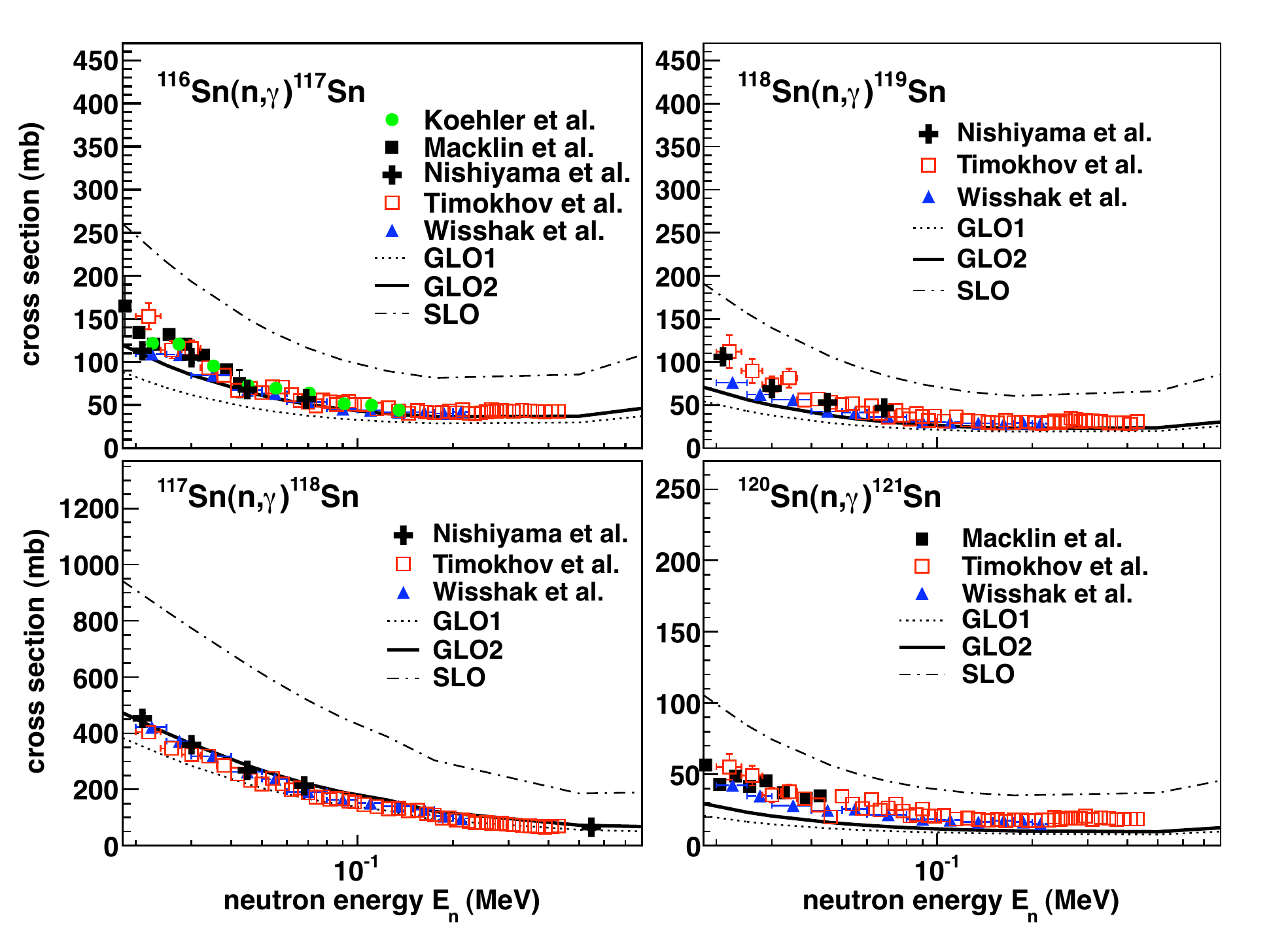}%[clip,width=1.7\columnwidth]
%\vskip 2cm
 \caption{(Color online) Data on neutron-capture cross sections for 
 the target nuclei $^{116-118,120}$Sn 
 compared with calculations, for neutron energies $\gtrsim 20$ keV. GLO1 (dotted line) is the GLO model assuming  constant temperatures
 (given in Tabs.~\ref{tab:pygme} and \ref{tab:parametre-teori116-119}), and GLO2 (solid line) is the GLO1 model plus the prediction of the pygmy resonance.  
 Measurements from Koehler \textit{et al.}~\cite{koehler}, Macklin \textit{et al.}~\cite{macklin}, Nishiyama \textit{et al.}~\cite{nishi}, Wisshak \textit{et al.}~\cite{wisshak}, and Timokhov \textit{et al.}~\cite{timokhov}.
 }
 \label{fig:allxsec}
 \end{center}
 \end{figure*}

The results of the comparisons
are shown in Fig.~\ref{fig:allxsec}. 
Our modeled strength function with a Gaussian pygmy resonance (denoted GLO2) leads to 
a calculated cross section that generally agrees very well with the measurements. Assuming the GLO model 
with constant temperature but without the pygmy resonance (GLO1), clearly gives a lower
cross section in all cases, as expected. 
This may be taken as a support of the finding of an enhanced strength function in the $E_\gamma$ region of $\approx 5-11$ MeV.
The SLO model gives an overall too high cross section,
which is not surprising considering the large overshoot in $\gamma$-ray strength compared to our measurements
and also to $\left< \Gamma_\gamma \right>$ data. We note that our calculated cross section for 
$^{116}$Sn$(n,\gamma)$$^{117}$Sn using the GLO2 model is in very good agreement with the one in the work of 
Utsunomiya \textit{et al.}~\cite{Utsunomiya}.

For the $^{117}$Sn$(n,\gamma)$$^{118}$Sn case, we have applied the model parameters that correspond to our scaled data (with a factor of 1.8). 
The resulting  excellent agreement with the experimental ($n,\gamma)$ data
further supports our choice of renormalization in Ref.~\cite{118-119}. We have in addition tested the strength prediction  using parameters that
fit with the original normalization, which results in  calculated cross section that are  clearly underpredictive compared to the experimental data (not shown here).

The reproduction of the $^{120}$Sn$(n,\gamma)$$^{121}$Sn cross section is not as good as for the
other nuclei, as this calculation seems to be somewhat more  underpredictive. However, the calculation is certainly an improvement compared to that of the GLO1, which is a standard  strength model  without the pygmy resonance.  
The underprediction might be explained by too low  experimental value of $\left< \Gamma_\gamma \right>$  in the normalization procedure of the measurements. If the value had been higher, the  pygmy resonance would had  produced more strength, leading to a general increase of the 
calculated cross section.  

\begin{figure}[!tb]
\includegraphics[width=9.5cm]{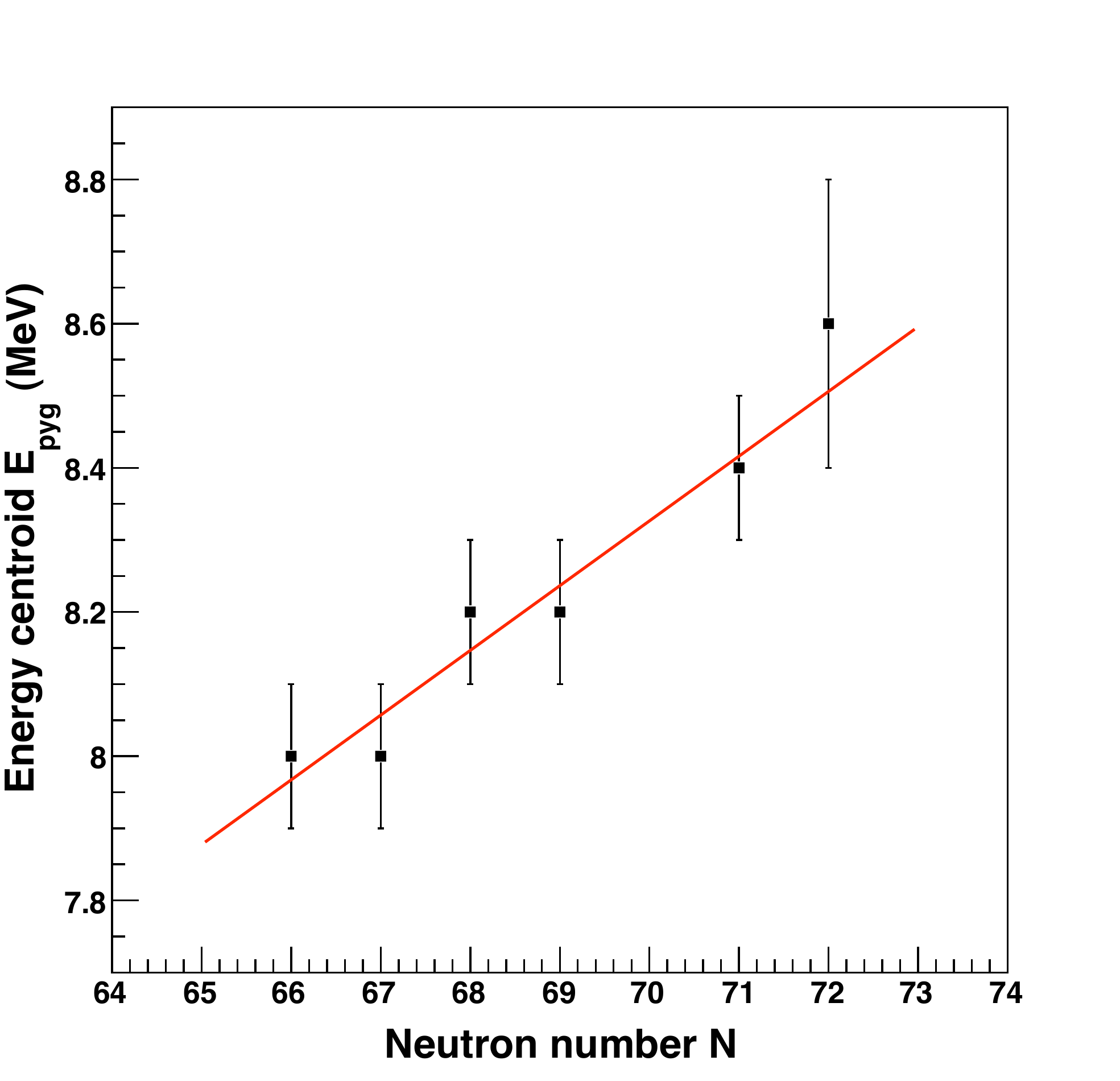}
\caption{(Color online) Estimated centroid energies $E_{\rm pyg}$ (squares) as a function of neutron number $N$, deduced from our  measurements on $^{116-119,121,122}$Sn.
The solid line is a linear $\chi^2$ fit to the measurements.}
\label{fig:energy-centroid}
\end{figure}

We would also like to study the evolution of the resonances' centroid energy $E_{\rm pyg}$
 with neutron number $N$. %of the resonance's centroid energies  $E_{\rm pyg}$.
Figure \ref{fig:energy-centroid} shows $E_{\rm pyg}$ as a function of  $N$ for the isotopes studied at the OCL. A $\chi^2$ fit has been performed on these data, resulting in the linear relation $E_{\rm pyg} = 2.0(16) + 0.090(23)\cdot N$ in units of MeV.

The estimates on $E_{\rm pyg}$ from others' experimental data on Sn are in agreement within the uncertainties with the observed pattern: 
$\approx 8.5$ MeV for $^{116,117}$Sn \cite{Utsunomiya},
 $\approx 7.8$ MeV for $^{117,119}$Sn \cite{Winhold}, and
10.1(7) and 9.8(7) MeV for $^{130,132}$Sn  \cite{Adrich}, respectively.
It is commented that  an increase of the resonances' centroid energies with increasing neutron number was also found in experimental data on Ca \cite{Tertychny}. 

The observation of an increase of the centroid energy with increasing neutron number   is not in agreement with common theoretical predictions.
On the contrary, studies on Sn isotopes  instead predict a  decrease of centroid energy with increasing neutron number. These studies include the
   Hartree-Fock-Bogoljubov (HFB) and  multiphonon quasiparticle-phonon (QPM) models by Tsoneva and Lenske \cite{Tsoneva},  the RHB+RQRPA (DD-ME2) model by Paar \cite{Paar07},  and
the Continuum QRPA model by Daoutidis \cite{Daoutidis}. 
Also a theoretical study on Ca isotopes,    using the Extended Theory of Finite Fermi Systems (ETFFS) by Tertychny \cite{Tertychny}, results in  a  decrease of centroid energy with  neutron number  (in contrary to experimental results on Ca, see Ref.~\cite{Tertychny} and references therein). However, it is commented that Daoutidis \cite{Daoutidis} predicts a relatively stable centroid energy in the atomic mass region $A=120-126$ compared to other mass regions. Hence, the  increase of centroid energy in the isotopes that we have compared, may be less than would had been observed in another mass region.

%This theoretical prediction is related to predominance  by   isoscalar over isovector states in the resonance.
%DETTE VAR OPPRINNELIG MED I ARTIKKELEN SOM BLE SENDT INN. SLETTES

%Our experimental observation of an increase of energy centroid for the more neutron-rich nuclei is a behavior one expects  from resonances of  predominantly isovector character. Thus, the isovector component might be more dominant  in the mixture of isoscalar and isovector states in the resonance than is often theoretically predicted. DETTE VAR OPPRINNELIG MED I ARTIKKELEN SOM BLE SENDT INN. SLETTES

Recent measurements using the ($\alpha,\alpha^\prime\gamma$) coincidence method on $^{124}$Sn compared
to photon-scattering experiments show a splitting into its isoscalar and isovector
components \cite{Endres2010}.
%Recent  measurements using the ($\alpha,\alpha^\prime\gamma$) coincidence method on $^{140}$Ce and $^{138}$Ba compared to NFR measurements  clearly show a splitting of the pygmy resonance into its isoscalar and isovector components  \cite{Endres}. 
Hence, both components seem to be present, in agreement with  theoretical predictions.

The nature, origin, and integrated strength of the Sn pygmy resonance are issues that are heavily debated. 
The  $E1$ neutron-skin oscillation mode, discussed in Refs.~\cite{Paar07, Sarchi, Isacker}, is   assumed as the underlying physical phenomenon in most of the theoretical predictions, both in  macroscopic (e.g., Van Isacker {\em et al.}~\cite{Isacker}) and microscopic approaches (e.g., Daoutidis \cite{Daoutidis}, Tsoneva and Lenske \cite{Tsoneva}, Paar \cite{Paar07}, Sarchi {\em et al.}~\cite{Sarchi}). Most theoretical calculations  predict 
 a systematic increase of the resonances' strength as the number of neutrons increase, due to the increase of the number of neutrons in the skin.  
 Another prediction is the increase by neutron number up to $^{120}$Sn followed by a decrease (Paar  \cite{Paar07}).
 Several of the   predicted increases of integrated strength concerning the isotopes that we have performed measurements on, are significant (e.g., Tsoneva and Lenske \cite{Tsoneva}, Van Isacker {\em et al.}~\cite{Isacker},  Litvinova {\em et al.}~\cite{Litvinova}). It is commented that the  study by Daoutidis \cite{Daoutidis}  predicts that also the integrated strength is relatively stable in the mass region  $A=120-126$. %compared to other mass regions. Hence, the  increase in integrated strength for the isotopes we have studied, may be  expected to be less than elsewhere.
%However, it is commented that the study of Daoutidis \cite{Daoutidis} predicted a relatively stabile energy centroid in especially the atomic mass region $A=120-126$, and that the increase of energy centroid in our isotopes may be  expected  to be less than elsewhere.

\begin{figure}[!htb]
\includegraphics[width=9.5cm]{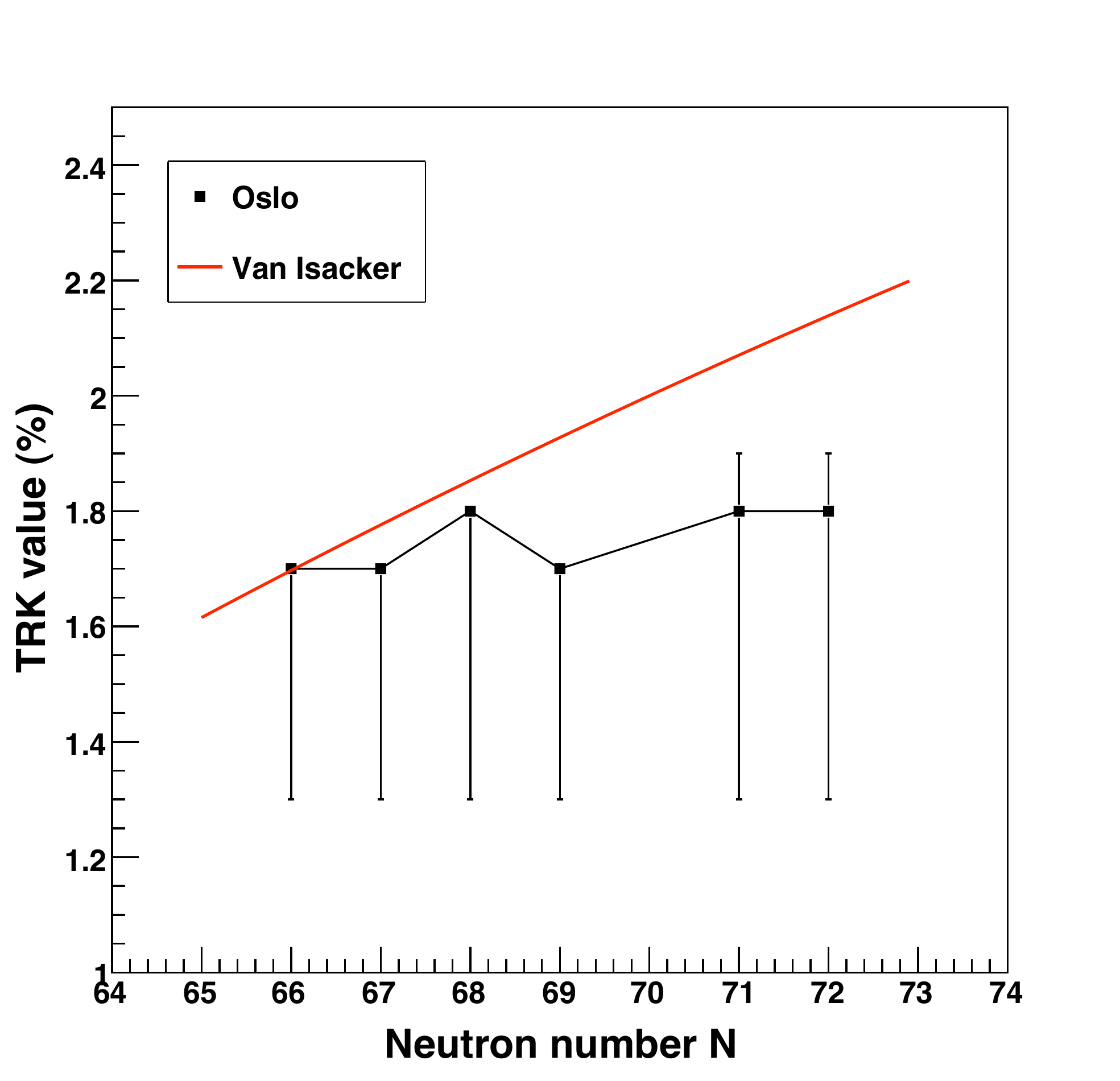}
\caption{(Color online) TRK values for Sn estimated from our measurements (squares) compared to the theoretical prediction from Van Isacker {\em et al.}~\cite{Isacker} (multiplied by a factor 14) (solid line) as a function of neutron number $N$.}
\label{fig:TRK}
\end{figure}

However for our measurements on the pygmy resonances in $^{116-119,121,122}$Sn, we cannot see any  dependency  on neutron number  in the integrated strengths. The same  resonance prediction has on the contrary been applied  for all isotopes.
(The total  integrated strengths and the TRK values of $^{121,122}$Sn being slightly larger than those of $^{116,117}$Sn, see Tabs.~\ref{tab:pygme} and \ref{tab:pygme}, is explained by differences in the GLO models of those isotopes.)
Figure \ref{fig:TRK} shows our TRK values together with those of Van Isacker {\em et al.}~\cite{Isacker} (multiplied by a factor of 14 in absolute value). The experimental result does not follow the predicted increase.
Still,  the uncertainties in our estimated resonance strengths are large. %, which is mainly due to the uncertainties in the measurements and the baseline. 
More experimental information is therefore needed in order to answer the question if the integrated  strength in Sn increases with   neutron number.

The experimental TRK values based on our and others' measurements are relatively large, compared to general excitations. This  might indicate that   the  pygmy resonance   is  due to a collective phenomenon. 
However, its origin is unknown, and single-particle excitations are not excluded.
Various theoretical predictions disagree on whether the neutron skin-oscillation is collective or not \cite{Paar07}. 

A clarification of the electromagnetic character  of the pygmy resonance in Sn would be of utmost importance.  
Present theoretical predictions assumes an $E1$ strength,  modeling the resonance  as a neutron-skin oscillation.
Experimental studies have  indicated an $E1$ character. Amongst these are the nuclear resonance fluorescence (NRF) experiments  performed on  $^{116,124}$Sn \cite{Govaert}   and on $^{112,124}$Sn \cite{Tonchev}, and the Coulomb dissociation  experiments performed on $^{129-132}$Sn \cite{Adrich,Klimkiewicz}. In addition comes the polarized photon beams experiments on $^{138}$Ba  \cite{TonchevPRL}. 
However, the existence of an $M1$ component of the resonance strength cannot be ruled out, as was discussed in Ref.~\cite{118-119}.

\section{Conclusions}  

The level density and  $\gamma$-ray strength functions of $^{121,122}$Sn have been measured using the ($^3$He,$^3$He$^\prime\gamma$) and ($^3$He,$\alpha\gamma$) reactions and the Oslo method.
The  level densities  of $^{121,122}$Sn display step-like structures for excitation energies below $\approx 4$ MeV. One of the bumps is interpreted as  a signature of  neutron pair breaking, in accordance with the findings in $^{116-119}$Sn.

A significant enhancement in the $\gamma$-ray strength  is observed in the $^{121,122}$Sn  measurements   for $E_\gamma \gtrsim 5.2$ MeV. The integrated strength of the resonances correspond to $\approx1.8^{+1}_{-5}\%$ of the TRK sum rule. These enhancements are compatible with pygmy resonances centered at $\approx$ 8.4(1) and $\approx 8.6(2)$ MeV, respectively.

Neutron-capture cross-section calculations using our  pygmy resonance  predictions give significantly better reproduction of   experimental  $(n,\gamma)$  cross sections than standard strength models without any  pygmy resonance.

The  pygmy resonances are  compared  to those  observed in $^{116-119}$Sn. 
The evolution with increasing neutron number of the pygmy resonances  observed in $^{116-119,121,122}$Sn is  a clear increase of  centroid energy  from 8.0(1) MeV in $^{116}$Sn to 8.6(2) MeV in $^{122}$Sn, while no  difference in integrated strength is observed. This finding is not in agreement with most theoretical predictions. However,
 the experimental uncertainties  are large, and more experimental information is needed in order to determine the nature of the pygmy resonances in the Sn 
 isotopes.

\acknowledgments
The authors would like to thank H.~Utsunomiya for giving us access to yet unpublished experimental results from photoneutron cross-section reactions 
$^{120,122}{\rm Sn}(\gamma,n)$.
The Department of Physics at the University of Jyväskylä (JYFL)  is thanked for kindly lending us the $^{122}$Sn target.
Last but not least, thanks to  
 E.~A.~Olsen, J.~Wikne, and A.~Semchenkov for excellent experimental conditions. 
The funding of this research from  The Research Council of Norway (Norges forskningsråd) is gratefully acknowledged.

\end{document}